%% file: main.tex
\documentclass[conference]{IEEEtran}
\IEEEoverridecommandlockouts
\usepackage{cite}
\usepackage{amsmath}
\usepackage{amssymb}
\usepackage{amsthm}
\usepackage{bbding}
\usepackage{graphicx}
\usepackage{textcomp}
\usepackage{xcolor}
\input{preamble}
\usepackage[nopostdot,nogroupskip,nonumberlist]{glossaries}
\usepackage[noend]{algpseudocode}
\usepackage{algorithmicx}
\usepackage{algorithm}
\usepackage{tikz,pgf}
\usepackage{pgfplotstable}
\usepackage{caption}
\usepackage{subcaption}
\usepackage{hhline}
\usepackage{booktabs}
\usepackage{footnote}
\usepackage{tablefootnote}
\usepackage{threeparttable}
\makesavenoteenv{tabular}
\makesavenoteenv{table}

\def\BibTeX{{\rm B\kern-.05em{\sc i\kern-.025em b}\kern-.08em
    T\kern-.1667em\lower.7ex\hbox{E}\kern-.125emX}}
\begin{document}

\title{\protocolname: Inclusive and Incentive-compatible\\ Vote Aggregation
%
\thanks{This work is partially funded by the BBChain and Credence projects under grants 274451 and 288126 from the Research Council of Norway.}
}

\author{\IEEEauthorblockN{Arian Baloochestani, Hanish Gogada, Leander Jehl, Hein Meling}
\IEEEauthorblockA{\textit{Department of Electrical Engineering and Computer Science} \\
\textit{University of Stavanger}\\
Stavanger, Norway \\
\{arian.masoudbaloochestani, hanish.gogada, leander.jehl, hein.meling\}@uis.no}}

\newtheorem*{example-non}{Example}

\algblockdefx{On}{EndOn}[1]{\textbf{on} #1}{\textbf{end on}}
\algblockdefx{Start}{EndStart}[1]{\textbf{start} #1}{\textbf{end start}}
\algblockdefx{Variables}{EndVariables}[1]{\textbf{Public Variables:} #1}{\textbf{end variables}}
\algblockdefx{PrVariables}{EndPrVariables}[1]{\textbf{Process Variables:} #1}{\textbf{end variables}}

\algnewcommand\algorithmicsend{\textbf{send}}
\algnewcommand\algorithmicto{\textbf{to}}
\algnewcommand\Send[2]{\State\algorithmicsend\ #1 \algorithmicto\ #2}

\algnewcommand\algorithmicverify{\textbf{verify}}
\algnewcommand\Verify[1]{\State\algorithmicverify\ #1}

\algnewcommand\algorithmicassert{\textbf{assert}}
\algnewcommand\Assert[1]{\State\algorithmicassert\ #1}

\algnewcommand\algorithmicstartt{\textbf{start}}
\algnewcommand\Startt[1]{\State\algorithmicstartt\ #1}

\algnewcommand\algorithmicreturnn{\textbf{return}}
\algnewcommand\Returnn[1]{\State\algorithmicreturnn\ #1}

\algnewcommand{\algorithmicand}{\textbf{ and }}
\algnewcommand{\algorithmicor}{\textbf{ or }}
\algnewcommand{\Or}{\algorithmicor}
\algnewcommand{\And}{\algorithmicand}

\makeatletter
\ifthenelse{\equal{\ALG@noend}{t}}%
  {\algtext*{EndOn}}
  {}%
\makeatother
\makeatletter
\ifthenelse{\equal{\ALG@noend}{t}}%
  {\algtext*{EndVariables}}
  {}%
\makeatother
\makeatletter
\ifthenelse{\equal{\ALG@noend}{t}}%
  {\algtext*{EndPrVariables}}
  {}%
\makeatother

\newacronym{QC}{QC}{quorum certificate}
\newacronym{PoW}{PoW}{proof-of-work}
\newacronym{BFT}{BFT}{Byzantine Fault Tolerance}

\newtheorem{theorem}{Theorem}
\newtheorem{lemma}{Lemma}
\newtheorem{corollary}{Corollary}
\newtheorem{remark}{Remark}

\maketitle
\thispagestyle{plain}
\pagestyle{plain}

\begin{abstract}
\input{tex/abstract.tex}
\end{abstract}

\begin{IEEEkeywords}
Committee-based blockchains, Vote omission attack, Vote inclusion, Signature aggregation, Incentive-compatible
\end{IEEEkeywords}

\section{Introduction}
\label{sec:intro}
\input{tex/intro}

\section{Background}
\label{sec:background}
\input{tex/background}

\section{System Model}
\label{sec:system}
\input{tex/system}

\section{Problem Statement}
\label{sec:problem}
\input{tex/problem}

\section{\protocolname}
\label{sec:model}
\input{tex/model.tex}

\section{Incentive Analysis}
\label{sec:incentives}
\input{tex/incentives}

\section{Security Analysis}
\label{sec:security}
\input{tex/security.tex}

\section{Experimental Results}
\label{sec:experiments}
\input{tex/experiments.tex}

\section{Conclusion}
\label{sec:conclusion}
\input{tex/conclusion.tex}

\clearpage 

\bibliographystyle{IEEEtran}
\bibliography{references}


\end{document}

%% file: preamble.tex
\usepackage{tikz}
\usetikzlibrary{shapes,arrows,chains,decorations.text}
\usepackage{siunitx}

\pgfdeclarelayer{marx}
\pgfsetlayers{main,marx}
\providecommand{\cmark}[2][]{%
  \begin{pgfonlayer}{marx}
    \node [nmark] at (c#2#1) {#2};
  \end{pgfonlayer}{marx}
  }
\providecommand{\cmark}[2][]{\relax}

\usepackage{todonotes}

\usepackage{acronym}
\acrodef{PoS}{Proof-of-Stake}
\acrodef{PoW}{Proof-of-Work}
\acrodef{VRF}{Verifiable Random Function}

\usepackage{xspace}
\newcommand{\protocolname}{Iniva\xspace}
\newcommand{\proj}{Iniva\xspace}

\usepackage{flushend}
\usepackage{titlecaps}

\usepackage{url}

\theoremstyle{definition}
\newtheorem{definition}{Definition}

\newcommand{\scndchance}{\textsc{2nd-chance}\xspace}

\newcommand{\novote}{vote denial\xspace}
\newcommand{\stealing}{aggregation denial\xspace}
\newcommand{\forced}{aggregation omission\xspace}

\newcommand{\sctimer}{\textit{secondChanceTimer}\xspace}
\newcommand{\aggtimer}{\textit{aggTimer}\xspace}
\newcommand{\sctimerl}{second chance timer\xspace}
\newcommand{\aggtimerl}{aggregation timer\xspace}

%% file: tex/abstract.tex
Many blockchain platforms use committee-based consensus for scalability, finality, and security.
In this consensus scheme, a committee decides which blocks get appended to the chain, typically through several voting phases. 
Platforms typically leverage the committee members' recorded votes to reward, punish, or detect failures. 
A common approach is to let the block proposer decide which votes to include, opening the door to possible attacks. 
For example, a malicious proposer can omit votes from targeted committee members, resulting in lost profits and, ultimately, their departure from the system.

This paper presents \proj, an inclusive and incentive-compatible vote aggregation scheme that prevents such vote omission attacks. 
\proj relies on a tree overlay with carefully selected fallback paths, making it robust against process failures without needing reconfiguration or additional redundancy. 
Our analysis shows that \proj significantly reduces the chance to omit individual votes while ensuring that omitting many votes incurs a significant cost. 
In addition, our experimental results show that \proj enjoys robustness, scalability, and reasonable throughput.

%% file: tex/intro.tex
\noindent
\sloppy
Recently, Ethereum~\cite{buterin2018ethereum}, the second largest permissionless blockchain system and the most popular smart-contract platform, completed its shift from \ac{PoW} to a \ac{PoS}-based consensus mechanism~\cite{themerge}. 
Similar to other networks, like Cosmos~\cite{kwon2016cosmos} or Algorand~\cite{chen2019algorand}, Ethereum now uses a committee-based consensus mechanism.
In committee-based consensus, a new block needs to be accepted and voted for by multiple processes from a committee.
Committee-based consensus can improve security and finality of \ac{PoS}~\cite{liu2020fairselection, amoussou2019fairnessCommittee}.
However, committee-based consensus creates new challenges, e.g., how to reward committee members.
To encourage participation and prevent free-riding, both Cosmos and Ethereum reward only active committee members~\cite{buterin2020incentives}.
Here, active committee members are detected through the inclusion of their signatures in the blockchain.
Therefore, it is crucial that the system includes all active members' signatures in fault-free cases. 

\sloppy
This reward scheme introduces the possibility of novel forms of attacks.
One such attack is the \textit{vote omission} attack, wherein a malicious actor or a colluding subset of the committee intentionally omits votes from a targeted victim.
This can drastically affect the victim's profitability and could even deter them from further participation in the system~\cite{buterin2018discouragement}.
While existing systems attempt to mitigate vote omission through carefully crafted incentive mechanisms~\cite{baloochestani2022rebop, kwon2016cosmos}, these strategies fail to address essential concerns.
Attacks can occur even when there is no immediate, discernible monetary gain for the attacker.
For instance, an attacker could strategically offset their losses through external mechanisms, such as short-selling on another platform.
Consequently, relying solely on monetary deterrents may be insufficient for preventing malicious activities like vote omission.
A more nuanced approach to incentives is crucial for enhancing the robustness of these systems.

\sloppy
Moreover, vote omission attacks are also feasible in permissioned systems without a reward mechanism.
For example, Carousel~\cite{cohen2022carousel} uses vote inclusion to select processes eligible for leadership.
Thus, a vote omission attack in this context may reduce the chances of electing a correct leader.

\sloppy
Addressing the issue of targeted vote omission is challenging.
Preventing omissions by individual processes requires redundant aggregation paths.
However, existing randomized approaches that use redundant paths allow free-riding.
Randomized approaches remain functional even when a large fraction of processes evade their aggregation duties, free-riding on others' work.
Vote aggregation, with its compute-intensive signature verification, is particularly attractive to avoid, especially if pairing-based signatures like BLS~\cite{bls12} are used.
Such free-riding again reduces redundancy in vote aggregation and thus simplifies vote omission.
Hence, we want aggregation protocols that are \textit{incentive-compatible}, meaning that processes face penalties or forfeit rewards if they neglect their aggregation responsibilities.

\sloppy
We analyze existing aggregation schemes.
Tree-based protocols like Kauri~\cite{neiheiser2021kauri} and ByzCoin~\cite{kogias2016byzcoin} lack the necessary redundancy to guard effectively against these attacks.
On the other hand, randomized approaches like Handel~\cite{begassat2019handel} and Gosig~\cite{li2020gosig} offer redundant aggregation paths but, ironically, this redundancy enables free-riding.
Our in-depth analysis shows that Gosig is only effective at mitigating vote omission under specific configurations.
Moreover, the very presence of free-riding exacerbates the potency of vote omission attacks.

\sloppy
This paper introduces \proj, a novel method to aggregate votes in committee-based blockchains.
Instead of relying on incentives, \proj leverages the properties of indivisible multi-signatures to effectively counteract vote omissions.
In common multi-signature schemes like BLS~\cite{bls12}, aggregated signatures cannot be decomposed into their constituent parts. 
Moreover, \proj organizes processes in a two-level tree.
With this structure, the root cannot omit individual votes aggregated at lower levels, while votes omitted at intermediate levels can be re-added.
This design effectively neutralizes targeted vote omissions.

\proj adopts a redundancy model based on fallback paths, activated only when required.
This approach strikes a balance, avoiding redundancy in fault-free scenarios while offering robustness against process and link failures.
By employing fallback paths, \proj eliminates the complex reconfiguration steps commonly used in other protocols~\cite{neiheiser2021kauri} to find a working tree.
Additionally, \proj's reward mechanism discourages free-riding during vote aggregation.
Since fallback paths are activated only under specific conditions, \proj can precisely determine which processes have fulfilled their aggregation duties.

\sloppy
We integrated \proj into the HotStuff consensus algorithm~\cite{yin2019hotstuff}.
Our experiments show that \proj ensures vote inclusion, even in the presence of faults. 
Additionally, \proj is scalable and has a reasonable performance overhead.
In summary, our key contributions are as follows:

\begin{itemize}
    \item We define \textit{indivisibility} as a property for multi-signature schemes, a property provided by existing aggregation schemes like BLS, and demonstrate its efficacy in mitigating targeted vote omission attacks.
    \item We present \proj, a robust vote aggregation and reward scheme for committee-based blockchains that significantly improves security against vote omission attacks. 
    \item We analyze our rewarding scheme using game theory, and prove its incentive-compatibility. 
    \item We analyze \proj's security and evaluate its effectiveness. Our analysis shows that for an attacker controlling 10\% of the processes, the chances to omit an individual signature are reduced by a factor of 10, while the cost of larger exclusion is increased by a factor of 7.
    \item We simulate vote omission attacks against Gosig and analyze the impact of free-riding. 
    \item We elaborate on the integration of \proj into the HotStuff protocol and conduct several experiments to analyze \proj's effectiveness in terms of scalability, throughput, latency, and vote inclusiveness. 
\end{itemize}

%% file: tex/background.tex
\subsection {Committee-based Blockchains}
\label{subsec:committee}
\input{tex/committee.tex}

\subsection {Multi-Signature Aggregation}

\noindent
Some committee-based blockchains elect one process as the leader to propose new blocks and receive all the votes to reduce the message complexity.
The scalability of these blockchains is dependent on the computational and network capacity of the leader~\cite{neiheiser2021kauri}.
Some previous works, such as HotStuff~\cite{yin2019hotstuff} rely on multi-signature aggregation schemes to reduce the message size by compacting all signatures into a single signature.
However, since HotStuff adopts a star topology, this puts even more load on the leaders by making them responsible for signature aggregation and sharing the result with all committee members.
Prior works have proposed decreasing the leader's load by distributing the aggregation work over some or all of the processes.
Kauri~\cite{neiheiser2021kauri} and ByzCoin~\cite{kogias2016byzcoin} use a tree overlay, where parents aggregate their children's votes.
Gosig~\cite{li2020gosig} uses a randomized overlay.
We discuss these approaches in more detail in the following.

\subsubsection{HotStuff}
\label{subsec:hotstuff}
\input{tex/hotstuff.tex}

\subsubsection{Tree-based approaches}
\label{subsec:kauri}
\input{tex/kauri.tex}

\subsubsection{Gosig}
\label{subsec:gosig}
\input{tex/gosig.tex}

%% file: tex/committee.tex
\noindent
\textit{Blockchain} is a list of blocks cryptographically linked to form a distributed ledger maintained and shared among all participants in a network.
Each block contains some data, e.g., transactions detail.
In addition, to ensure the integrity of the blockchain, each block also contains the hash of its previous block.
Hence, once a block is added to the blockchain, it is considered immutable since any modification to the block would also change its hash.
Each process in the network holds a public/private key pair, and their identities are verified through digital signatures.

Processes need to follow a \textit{consensus algorithm} to agree on the inclusion of blocks into the chain.
Bitcoin~\cite{nakamoto2008bitcoin} introduced the \ac{PoW} consensus algorithm.
While \ac{PoW} guarantees security, it suffers from several drawbacks, such as probabilistic consistency (forks) and high computational overhead~\cite{meng2018committee}.
To this end, some blockchains~\cite{kwon2014tendermint,chen2019algorand,baudet2019LibraBFT,daian2019snow,hanke2018dfinity} adopt classical \gls{BFT} protocols~\cite{castro1999practical,lamport2019byzantine} as the consensus algorithm.
However, as these protocols do not scale to a large number of processes, these methods use a small committee to run the consensus algorithm. 

In committee-based blockchains, first, a leader is elected to propose a block. 
Then, a selected committee verifies the block and votes by signing the block using digital signature schemes.
Leaders gather the signed blocks, and if a block gains more than a fraction of the votes, the block is considered approved.

Designing a fair rewarding mechanism for committee-based blockchains is challenging. 
To prevent free riding, only active members should get rewarded~\cite{amoussou2020rational}.
Most current protocols rely on the leaders to detect the active members by collecting the list of voters. 
As the leaders might deviate from the protocol, existing methods incentivize them to act correctly. 
Cosmos~\cite{kwon2016cosmos} introduces the \textit{variational bonus} mechanism in which leaders receive an extra fraction of the reward based on the number of votes they collect from the previous committee.
Rebop~\cite{baloochestani2022rebop} proposes a reputation-based leader election mechanism with the reputation defined as the number of collected votes in the last $T$ rounds as the leader.

%% file: tex/hotstuff.tex
HotStuff is a consensus protocol that operates in a sequence of views, each involving three voting rounds.
The first round is the \textit{prepare} round, where the leader proposes block $B$ for view $v$ and height $h$.
The committee members then validate $B$ and vote with a \textit{prepare} message if they have not already prepared a block with a higher view at the same height.
Once the leader receives enough \textit{prepare} votes, it aggregates them into one signature called a \gls{QC} and shares it with all processes.
Processes record the received \gls{QC} and respond with a \textit{pre-commit} message.
The leader waits for enough \textit{pre-commit} replies, forming another \gls{QC}.
This \gls{QC} is then sent to all processes in the final \textit{commit} round, resulting in a block commitment once enough commit votes are received.
According to Yin et al~\cite{hotstuffArxiv}, protocols like Casper~FFG~\cite{buterin2017casper} used in Ethereum, or Tendermint~\cite{kwon2014tendermint}
used in Cosmos can be seen as variants of HotStuff.

To achieve better performance, the three rounds can be performed concurrently for three different views.
This variant is called chained HotStuff, where a single \gls{QC} can serve as \textit{prepareQC}, \textit{pre-commitQC}, and \textit{commitQC} at the same time.

HotStuff, can use different leader election policies.
Blockchains typically adopt the Leader-Speak-Once (LSO) model~\cite{abraham2021rotating, giridharan2023beegees}, where every leader only proposes a single block and the leader is changed every view.
LSO minimizes the leader's power over new block proposals and makes the protocol more fair.

%% file: tex/kauri.tex
Kauri and ByzCoin use a tree for distributing the signature aggregation work among the processes. 
The tree-based topology reduces the workload on the leaders compared to the HotStuff star topology because each parent process is responsible for aggregating its sub-tree.  
In case of failure, these protocols require reconfigurations and may fall back to a star topology in cases with many failures~\cite{neiheiser2021kauri}.  
Kauri proposes a reconfiguration mechanism for trees with height 2. 
In these trees, processes need to aggregate $\mathcal{O}(\sqrt{n})$ many signatures.
Kauri uses pipelining techniques to achieve high throughput despite the added latency through communication on the tree.
However, while the aggregation work is distributed among the processes, the parent processes have complete control over their sub-tree and are able to exclude leaf children from the aggregated signature.

%% file: tex/gosig.tex
Gosig~\cite{li2020gosig} is a BFT protocol for committee-based blockchains.
In Gosig, leaders are selected secretly using a \ac{VRF} and share their block proposals with other processes.
Each process performs signature aggregation and repeatedly shares its current aggregate with $k$ other processes, selected at random from the complete committee.

%% file: tex/system.tex
\noindent
A set $\Pi = \{p_1,p_2,...,p_n\}$ of processes are available in the committee.
For simplicity, we assume $\Pi$ to be constant and do not consider the committee selection protocols.
The fixed membership assumption is to simplify explanation and analysis.
Our solution also works for dynamic committees as long as committee members for one view are known a priori.
We assume a synchronous network with an upper bound $\Delta$ on the delivery of any message between correct participants.
Our system requires synchrony to ensure inclusiveness.
In the case of an eventually synchronous system, it ensures inclusiveness after global stabilization time~\cite{gst}.

We assume an adversary controlling a fraction $m$ of the processes in the committee, where $m\leq f =1/3$.
Processes under the control of the adversary may behave arbitrarily.
However, we are especially interested in the case where the adversary tries to diminish the reward received by one victim $p_v\in\Pi$.
We assume that the adversary cannot disturb the processing and communication between correct processes.
Thus, denial of service attacks are out of scope.

We assume each process $p_i$ in the system has a private/public key pair $sk_i/pk_i$ and access to a list of other processes with their public keys.

A multi-signature scheme is a digital signature scheme that allows the aggregation of signatures.
Let $\sigma_1=\text{sign}(m,sk_1)$ and $\sigma_2=\text{sign}(m,sk_2)$ be signatures for a message $m$ produced with different private keys.
The signatures can be aggregated with multiplicity $i$ and $j$ where $i,j \in \mathbb{Z}$:
$$\sigma'=\text{agg}(\sigma_1^i,\sigma_2^j)$$
The resulting signature $\sigma'$ can be verified by aggregating the corresponding public signatures with the same multiplicity: $$\text{verify}(\sigma',pk_1^i pk_2^j)$$

\noindent
We assume processes have access to an \textit{indivisible multi-signature} scheme, such that given $\sigma'$, it is infeasible to retrieve $\sigma_1$ or $\sigma_2$.
For pairing-based signatures, indivisibility of up to $k$ signatures was proposed as an assumption by Boneh et al~\cite{boneh2003aggregate}.
We use BLS signatures~\cite{boneh2004short}, which are indivisible according to Coron and Naccache~\cite{coron2003boneh}.

At the beginning of each round, $p_i$ is assigned a unique ID $(ID[p_i]= i)$.
We assume the processes have access to a deterministic shuffling algorithm, and $\Pi$ is shuffled every round so that the IDs will be different at each round of the protocol.
The shuffling algorithm needs to be unpredictable, meaning that the processes cannot predict the outcome of the shuffling for future rounds.
As an example, the above algorithm can be implemented using a \ac{VRF}~\cite{micali1999verifiable}.

%% file: tex/problem.tex
\noindent
In committee-based blockchains a leader disseminates a block to participants, who return votes/signatures to the leader.
The leader then outputs an aggregate of these votes, aka a \gls{QC}.
We present a slight variation of Kauri's~\cite{neiheiser2021kauri} vote aggregation scheme below.

\newcommand{\blk}{\ensuremath{B}\xspace}

\begin{definition}
    A \textit{vote aggregation scheme} has an interface with the following communication primitives:
    \begin{itemize}
        \item\textbf{broadcast}(\blk). Invoked by the leader to disseminate a block \blk and start vote aggregation.
        \item Upcall \textit{deliver}(\blk) at $p_i$ delivers \blk. $p_i$ emits a vote for block \blk:
        \[
            \textit{vote}({\blk}) =
            \begin{cases}
            \sigma_{{\blk},i} & \text{if \blk is valid} \\
            \bot         & \text{if \blk is invalid}
            \end{cases}
        \]
        \item Upcall $\textit{aggregate}({\blk}, QC_{\blk}, md)$ at the leader delivers an aggregate $QC_{\blk}$ of valid signatures from \textit{vote}  and additional metadata $md$ specifying which processes' votes are included.
    \end{itemize}
\end{definition}

\noindent
Neiheiser et~al~\cite{neiheiser2021kauri} define the following liveness properties for a vote aggregation scheme:
\begin{definition}[\textit{Reliable Dissemination}]
    If the leader is correct, all correct processes deliver the block sent by the leader.
    \label{def:reliableconfig}
\end{definition}

\begin{definition}[\textit{Fulfillment}]
    If the leader is correct and all correct processes invoke $\text{vote}$ with a valid signature, then the leader emits a \gls{QC} containing at least $(1-f)N$ signatures.
    \label{def:fulfilment}
\end{definition}

\noindent
These properties are sufficient to ensure liveness and safety of HotStuff~\cite{neiheiser2021kauri}.
Additionally, straightforward validity properties are expected, i.e. that correct processes only deliver blocks actually sent by the leader, and that $QC_{\blk}$ only includes valid signatures.

In this work we are interested in the LSO model, where the leader changes after every block.
We therefore adapt the vote aggregation scheme, assuming that \textbf{broadcast} is invoked by the leader proposing \blk, while $\textit{aggregate}$ happens at the next leader.
Further, we require reliable dissemination and fulfillment to hold only if two consecutive leaders are correct.

\subsection{Rewarding}

\noindent
Some committee-based cryptocurrencies use the \gls{QC} to reward participants.
For example in cryptocurrencies like Cosmos~\cite{kwon2016cosmos}, Solidus~\cite{abraham2016solidus}, or Ethereum~\cite{buterin2018ethereum}, the \gls{QC} is used to detect active committee members and reward them accordingly to prevent free riding.

Such rewarding schemes can be modelled as a function $\textit{reward}(QC)$, which computes a distribution of rewards based on the quorum certificate.
Since the $QC$ is included in the next block, the reward distribution can be verified by every process, re-computing the $\textit{reward}$ function.

\noindent
\textbf{Inclusiveness:}
If the $QC$ is used for rewarding, it is crucial for these methods to guarantee the inclusion of all non-faulty processes within the \gls{QC}.
We refer to this attribute as being \textit{inclusive}.

\begin{definition}[\textit{Inclusiveness}]
    If the current and next leader are correct, then
    all signatures from correct processes are contained in the aggregated QC.
\label{def:inclusive}
\end{definition}

\noindent
We note that Inclusiveness may also be useful in other contexts.
For example, Carousel~\cite{cohen2022carousel} proposes a reputation-based leader rotation mechanism that looks at the previous \gls{QC}s to avoid selecting failed processes as leaders.
Using Carousel, inclusiveness can guarantee that all correct processes actually can become the leader.

\subsection{Vote omission}
\noindent
Since leaders are in charge of forming \gls{QC}s,
a malicious leader can ignore some of the votes and form the \gls{QC} with the processes it desires.
We refer to this attack as the \textit{vote omission} attack.
Incentive engineering~\cite{baloochestani2022rebop} can ensure vote omissions are not profitable.
However, attacks are still possible. Especially attacks targeted at an individual process may have a devastating effect on the victim, while only incuring a small cost to the attacker.
In \textit{targeted vote omission}, an attacker controlling a large fraction of the committee tries to omit as many votes from a specific process as possible.
In these attacks, the attacker does not intentionally omit other processes unless it leads to a more successful attack.
We define \textit{collateral} as the number of non-target processes that an attacker is willing to exclude to perform the attack.
For example, with a collateral of 0 only the target will be excluded and no other processes.
To measure the robustness of a protocol against targeted vote omission attacks, we define \textit{$c$-omission probability}.

\begin{definition}[\textit{$c$-omission probability}]
We define the $c$-omission probability as the probability for an attacker to successfully perform a targeted vote omission attack with collateral at most $c$ during one instance of vote aggregation based on a random assignment of processes to the attacker and the victim role.
The probability space is the set of all possible process assignments.
We assume all such assignments to be equally likely.
Omission probability is a function in $m \in [0,1]$, the fraction of the committee's processes controlled by the attacker.
\label{def:omission}
\end{definition}

\noindent
For instance, the HotStuff protocol adopts a round-robin leader selection scheme.
Thus, an attacker controlling a fraction $m$ of the processes can become the leader $m$ fraction of the time.
Given that each leader has the authority to decide which votes to incorporate, the probability of the attacker executing the targeted vote omission attack is $m$.

We note that as an attack probability, a $c$-omission probability of $m^2$ signifies a more robust protocol, than $c$-omission probability of $m$.

\subsection{Free riding}
\noindent
Vote aggregation schemes that support redundant aggregation are susceptible to free riding.
Free riding by other processes (neither victim, nor attacker) helps an attacker to perform vote omission.
For example, in Gosig, all processes are expected to participate in vote aggregation.
However, some processes may decide to omit the aggregation step to avoid  costly signature verification, and instead, only disseminate their own signature.
If other processes follow this \textit{free riding} behavior, it simplifies a targeted vote omission of correct processes, as our simulations show (see Section~\ref{sec:security}).
To avoid such free riding, we require vote aggregation to be \textit{incentive compatible}.

\begin{definition}[\textit{Incentive compatibility}]
A rewarding scheme is incentive compatible if following the protocol gives higher utility compare to other strategies.
\label{def:incentive}
\end{definition}

\subsection{Alternative approaches}
\noindent
While existing approaches for signature aggregation also use indivisible multi-signatures, they have multiple shortcomings.
A summary of the existing protocols' drawbacks is shown in Table~\ref{tab:agg-comparison}.

\input{./tables/table-comparison.tex}

Existing tree-based signature aggregation approaches such as Kauri or ByzCoin fail to prevent vote omission attacks as the internal processes in the tree have direct control over their children and are able to selectively omit them.
Both Kauri and ByzCoin use a stable tree whose reconfiguration is triggered by the leader.
This allows an attacker in charge of the leader to arrange a configuration where it also controls the parent of the victim.
Additionally, the failure of internal processes leads to the loss of the whole sub-tree under them.
This can result in omissions even in the absence of attacks since these methods are not inclusive.
Complex reconfiguration is needed in case of failures to rearrange the tree.

Gosig uses a randomized, redundant communication pattern for vote aggregation.
The inclusion of a given process in the \gls{QC} is therefore probabilistic, even in fault-free cases.
Here, if the attacker receives the victim's individual signature early in the aggregation process, it will be able to remove it from the final certificate.
We performed simulations on the omission probability of Gosig, which shows that it can reduce targeted vote omissions only for small  values of $k$ and attackers controlling only a small fraction $m$.
For larger values, Gosig $0$-omission probability is $m$,
allowing targeted omission every time the attacker is selected as leader.
Additionally, Gosig is vulnerable to free-riding, which simplifies targeted vote omission.

Another approach to reduce vote omission is to let processes compete in aggregation and use the process aggregating the most signatures as the next leader.
A similar approach was applied in Rebop~\cite{baloochestani2022rebop}.
Unfortunately, this approach opens novel attacks.
An attacker may hold back its own signature, thus reducing others' chances of leadership.
Note that as incentive engineering and reputation-based schemes such as Rebop~\cite{baloochestani2022rebop} can defend against targeted vote omission attacks with large collateral, we are mostly interested in collateral of $0$.

In the next section, we show how \protocolname avoids reconfiguration and omission using a tree-based overlay and its extension with an incentive scheme that prevents free riding.

%% file: tables/table-comparison.tex
\begin{table}
    \centering
    \resizebox{\linewidth}{!}{%
        \begin{threeparttable}
            \renewcommand{\arraystretch}{1.2}
            \caption{A comparison between existing multi-signature aggregated schemes}
            \label{tab:agg-comparison}
            \begin{tabular}{l|ccc}
                \toprule
                \multicolumn{1}{l|}{} & \textbf{$0$-omission probability} & \textbf{Inclusive} & \textbf{Incentive compatible} \\
                \midrule
                \textbf{Star protocol } & $m$ & Yes & Yes \\
                \textbf{Randomized tree } & $m$\tnote{a} & No & Yes \\
                \textbf{Gosig ($k$)} & $k$-dependent\tnote{b} & No & No \\
                \textbf{\protocolname} & $m^2 $ & Yes & Yes \\
                \bottomrule
            \end{tabular}
            \begin{tablenotes}
                \footnotesize
                \item[a] In a static configuration, the leader may perform the attack every round.
                \item[b] The $0$-omission probability of Gosig depends on $k$. See Section~\ref{sec:security}.
            \end{tablenotes}
        \end{threeparttable}%
    }
\end{table}

%% file: tex/model.tex
\input{figures/protocoloverview.tex}

\noindent
In committee-based blockchains, committee members work together to append a new block to the blockchain through several views. 
The current length of the blockchain is represented through the parameter height $h$. 
Processes move to the next view if they fail to append a new block, while height remains unchanged. 
At each view $v$, one of the processes is selected as the leader ($L_v \in \Pi$) and is responsible for proposing a new block. 
For adding the proposed block to the blockchain, $L_v$ must gather at least $1-f$ fraction of the votes from the previous committee, where $f$ defines the maximum fraction of faulty processes that the protocol can handle (e.g., $f=1/3$). 
An aggregated signature of $1-f$ fraction of the committee is called a \gls{QC}. 
The \gls{QC} of the last approved block is called the highest \gls{QC}. 
$L_v$ uses the highest \gls{QC} to distribute a reward $R$ among the members whose votes are included. 

In this section, we present \proj, an Inclusive and Incentive Compatible Vote Aggregation mechanism in committee-based blockchains. 
In the following, we first discuss the proposal propagation and vote aggregation in \proj, and then we present a rewarding scheme that makes \proj incentive compatible. 

\input{./algorithms/aggregation.tex}

\subsection{Signature Aggregation}
\noindent
In this section we discuss the block propagation and signature aggregation procedures in \protocolname, which are shown in Algorithm~\ref{alg:sig} and Figure~\ref{fig:protocol_overview}.

At the start of each view $v$, the leader of that view, $L_v$ creates a new block extending the blockchain at current height $h$, $B_{h+1}$. 
Based on the \gls{QC} and view number included in the block, all processes generate the same tree for the given view (Lines \ref{line:startview}-\ref{line:maketree}, Line \ref{line:maketree2}).
$L_v$ then forwards the block to the root process in the tree and its children (Line \ref{line:sendtonextleader}, Figure~\ref{fig:protocol_overview}-A).
After receiving and verifying a block, a process builds the tree itself and forwards the block to its children.
Processes without children (tree leaves) instead send their signatures to their parents (Lines \ref{line:onprop}-\ref{line:sendsign}). 

Each internal process in the tree verifies and aggregates the received signatures together with its own signature (Lines~\ref{line:onsig}- \ref{line:aggregate}). 
Upon a timeout, or once aggregation for all children is completed, the process forwards the aggregated signature to its parent (Line~\ref{line:send!leader}).
It also sends an acknowledgement (\textit{ack}) to its children (Line~\ref{line:sendack}). 
The ack includes the aggregated signature and acts as proof that the parent has included the signatures of the senders.  

Due to network issues or malicious processes in the tree, some processes may not receive the proposal and aggregated signatures may be incomplete.
The root process in the tree is the leader of the next view $L_{v+1}$.
The root process collects the signatures to a \gls{QC}, which it uses to create the next block. 
Before creating the next block, $L_{v+1}$ gives one last chance to the processes whose votes are not included by sending them a \scndchance message.
$L_{v+1}$ does send this message either once a QC has been collected or upon a timeout (Lines~\ref{line:leader}-\ref{line:lastchance}).

Replying to a \scndchance message with their individual signature enables the message sender to exclude a process. 
Therefore, processes reply to a \scndchance with the aggregated signature received from their parent in an \textit{ack} message.
Otherwise, \scndchance messages are validated according to function isValid. 
A second chance message is valid if it includes a quorum of signatures, or a signature from the parent, but not the current process's signature.
Additionally, a second chance message may also be valid if sufficient time has passed since the block creation.
This can be checked by comparing the block timestamp against the current time.

Since the internal tree processes do more work than other processes, we propose a mechanism to reward them for their extra work. 

\subsection{Rewarding Mechanism}
\noindent
We now explain our rewarding mechanism. Rewards are distributed by the leader or root. We first explain how rewards are distributed and then how other processes verify the distribution determined by the leader.
We identify the following requirements for our rewarding system:
\begin{enumerate}
    \item\label{req:vote} All active committee members should be rewarded.
    \item\label{req:extra} Processes with extra responsibilities, like the internal processes and the leader, should receive an additional reward.
    \item\label{req:freeRiding} Omission of any assigned duties, i.e. voting, aggregation, or \scndchance messages, should result in reduced rewards.
    \item\label{req:constant} The total reward paid out per block should be independent of how many votes were aggregated.
\end{enumerate}

\noindent
We note that requirements~\ref{req:vote}-\ref{req:freeRiding} ensure that processes are motivated to conduct their assigned tasks. 
Requirement~\ref{req:constant} ensures that the aggregation and rewarding procedures do not affect the amount being distributed.
This allows, for example, to use fees received from users to be redistributed as a reward. In case rewards are newly minted tokens, this ensures a constant and predictable creation rate.
Finally, this also ensures that our rewarding method is not susceptible to attacks, where a process may forfeit some of its rewards but receives a larger fraction of the total reward paid. Such attacks exist in other schemes, e.g. selfish mining~\cite{selfishmining}.

According to Requirement~\ref{req:extra} and \ref{req:vote}, we use a certain fraction of the total reward to give a bonus for aggregating processes ($b_a)$, and the leader ($b_l)$ and distribute the remaining reward evenly among all processes, whose signature is included in the final vote $b_v=(1-b_l-b_a)$.

Let $R$ denote the total reward given out for one block.
Due to Requirement~\ref{req:constant}, the bonus for aggregation and leader is given as a fraction of $R$. 
As a bonus for aggregation, internal processes receive $\frac{b_a}{n}R$ for each signature of a child.
Similarly, the leader, or root of the tree, receives $\frac{b_a}{n}R$ for each subtree that it aggregates.

For the leader bonus, we use a similar approach as the variational bonus introduced in Cosmos~\cite{kwon2016cosmos},
where the leader receives a bonus of $\frac{b_l}{fN}R$ for each signature included in the final certificate, exceeding the minimal requirement of $(1-f)N$ signatures.

The reason for having a separate bonus for the leader is that the leader is the only process that can send \scndchance messages to every other process. 
Therefore, by tying the leader bonus to the number of included processes, we motivate the leader to send \scndchance messages to all missing processes. 

Finally, we want leaf processes to be aggregated by their parents rather than through \scndchance messages. 
If a leaf process is included via a \scndchance message, its parent loses the $\frac{b_a}{n}R$ aggregation bonus.
In these cases, we also reduce the voting reward received by the child by $\frac{b_a}{n}R$.

Finally, all remaining reward, after deducing aggregation and leader bonuses and applying punishment for \scndchance, is distributed evenly among all the processes in the committee.

We note that to compute the rewards, it is necessary to know who the leader was, which signatures have been included, who performed how many aggregations, and whether signatures have been collected through aggregation or via \scndchance messages.

Since the leader and tree can be recreated deterministically, 
the main issue is determining if a signature has been collected through \scndchance messages.
For this purpose, we use the fact that the same signatures can also be aggregated multiple times in an indivisible aggregation scheme.
Thus, when an internal process aggregates its children, it includes each child's signature twice, while a leader aggregating \scndchance messages will include signatures only once. 
Additionally, the internal process will include its own signature one additional time for each aggregated child.

For example, if a process collects 2 signatures $\sigma_1$ and $\sigma_2$, it adds its own signature $\sigma_i$ 2 additional times, resulting in an aggregated signature:
\begin{equation}
\label{eq:example}
\aggsig=\text{agg}(\sigma_1^2,\sigma_2^2,\sigma_i^3)
\end{equation}

\noindent
The leader does check these multiplicities and only includes correctly aggregated shares.
We note that if an internal process or a leaf sets a wrong multiplicity on its signature, this can be detected by the leader.
Further, the leader cannot change the multiplicity of signatures reported by internal processes since these are indivisible.
To check that aggregation bonuses and \scndchance punishments are computed correctly, processes simply compare the multiplicities of the signatures of leaf and internal processes. 
The leader is considered faulty if the multiplicities reported in a block are wrong.

\subsection{Discussion}
\noindent
\proj uses a tree-based structure and indivisible multi-signature aggregation scheme to remain inclusive and prevent vote omission attacks.
In the absence of failures and attacks, \proj requires only one tree aggregation, which is comparable to existing tree-based aggregation schemes~\cite{neiheiser2021kauri} in terms of latency and throughput.
In the presence of partial failures, \proj relies on fallback paths for fault tolerance. 

\begin{theorem}
\label{the:reliablediss}
Algorithm~\ref{alg:sig} guarantees Reliable Dissemination. 
\end{theorem}

\begin{proof}
According to Definition~\ref{def:reliableconfig} and our adjustment to LSO, we assume that the leader $L_{v}$ and the next leader $L_{v+1}$ are correct. $L_{v+1}$ is also the root of the tree used for dissemination.
If any correct process $p_i$ does not receive the block through the tree dissemination (Line~\ref{line:deliver1} of Alg.~\ref{alg:sig}), $p_i$ will not send a signature.
Therefore $L_{v+1}$ will send a \scndchance message to $p_i$ and $p_i$ will deliver executing (Line~\ref{line:deliver2}).
\end{proof}

\noindent
In \proj we use a tree of height $2$ (Algorithm~\ref{alg:sig}).
A tree with more levels could provide better protection against vote omission, as the internal processes would also send \scndchance messages.
However, multiple rounds of \scndchance messages, and additional levels would significantly increase latency.

\proj's maximum latency for each round is $7 \Delta$.
Since $\Delta$ is the upper bound for message delivery between correct processes, it takes $1 \Delta$ for $L_{v-1}$ to share a new block with $L_{v}$.
Thus, leaf processes receive the block $2 \Delta$ later, and it takes another $2 \Delta$ for the leader to receive the aggregated messages.
Finally, if there are any missing signatures, another $2 \Delta$ is added to the overall latency due to the \scndchance messages.

\begin{theorem}
\label{the:inclusiveness}
Algorithm~\ref{alg:sig} guarantees Inclusiveness after $7\Delta$. 
\end{theorem}
\begin{proof}
Let $p_i$ be a correct process whose signature was not received by the root during tree aggregation.
Since we can assume that the root and next leader is correct, $p_i$ will receive a \scndchance message and reply either with its own signature, or an aggregate received in \textsc{ack}.
In the later case, the aggregate also includes $p_i$'s signature.
This signature will be aggregated by the leader.
The delay of $7\Delta$ follows from the argument above.
\end{proof}

\noindent
The following Corollary follows easily, since Inclusiveness actually implies Fulfillment.

\begin{corollary}
\label{cor:corr1}
Iniva guarantees Fulfillment.
\end{corollary}

\noindent
Note that the number of included votes is also dependent on when the leader send the \scndchance messages.
Processes that have not received the block from their parents need some time to verify and sign the block.
If the leader sends the \scndchance within a certain timeout, missing processes have more time to keep up.
However, some processes might receive the \scndchance message before the acknowledgment from their parent.
While increasing timeouts alleviates this problem, it leads to higher latency and lower throughput.
Our evaluations (section~\ref{sec:experiments}) show that in presence of failures, lower timeouts result in increased throughput, while larger timeouts favor inclusiveness.

%% file: figures/protocoloverview.tex
\begin{figure}[tb]
\includegraphics[width=\linewidth]{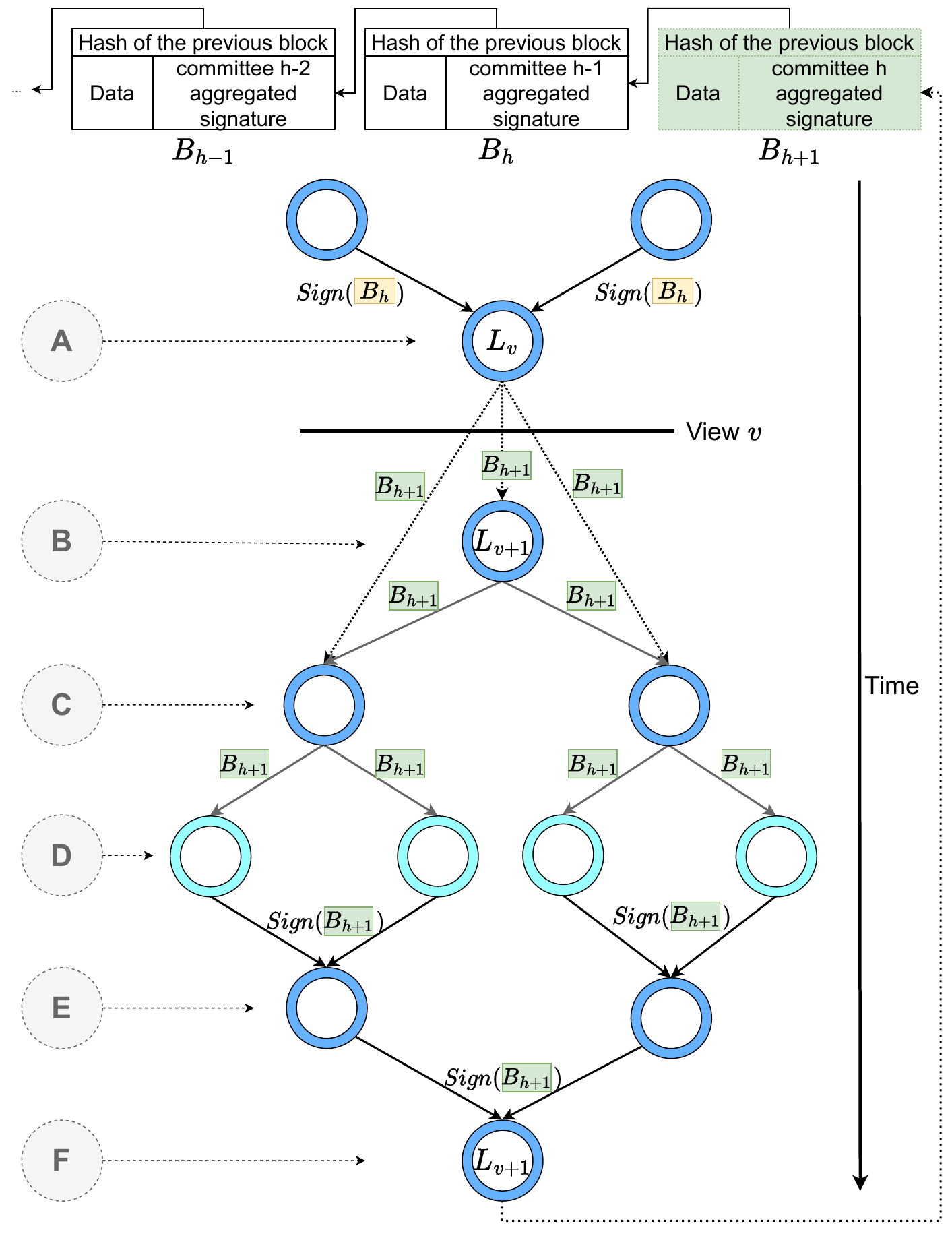}
\caption{An overview of \protocolname.
\textbf{A)}~$L_{v}$ commits $B_{h}$.
It creates and forwards $B_{h+1}$ to $L_{v+1}$ and $L_{v+1}$ children.
\textbf{B)}~$L_{v+1}$ receives $B_{h+1}$ and starts the view by sharing the proposal with its children.
\textbf{C)}~Internal nodes forward $B_{h+1}$ to their children, and wait for their response.
\textbf{D)}~Leaf nodes verify and sign  $B_{h+1}$, and share their signature with their parent.
\textbf{E)}~Internal nodes aggregate their children signatures, and share it with their parent.
\textbf{F)}~$L_{v+1}$ commits $B_{h+1}$. It creates and forwards $B_{h+2}$ to $L_{v+2}$ and $L_{v+2}$ children.}
\label{fig:protocol_overview}
\end{figure}

%% file: algorithms/aggregation.tex
\newcommand{\aggsig}{\textit{aggSig}}

\begin{algorithm}[t]
    \caption{Block propagation and signature aggregation}
    \label{alg:sig}
    \begin{algorithmic}[1] 
        
        \PrVariables
            \State \textit{parent} \Comment{Direct parent of the process in the tree}
            \State \aggsig \Comment{The aggregated signature}
        \EndPrVariables

        \vspace{0.1cm}

        \On \textnormal{\textbf{broadcast}}$(B)$
            \label{line:startview}
            \Comment{at leader $L_v$}
            \State{$\textit{root}, \textit{children} \gets \textnormal{makeTree}(B)$}
            \label{line:maketree}
            \Send{$\langle \textsc{proposal} , B \rangle$}
            {\textit{root} \text{and} \textit{children}}
            \label{line:sendtonextleader}
        \EndOn
        
        \vspace{0.1cm}
            \On $\langle \textsc{proposal} , B \rangle$
                \label{line:onprop}
                \State \textit{parent, children} $\gets \text{makeTree} (B)$ \label{line:maketree2}
                \If {$\textit{children} \neq \emptyset $}
                    \Send{$\langle \textsc{proposal} , B \rangle$}{\textit{children}}
                \EndIf
                \State $\textit{deliver}(B)$
                \label{line:deliver1}
                \State $\sigma_{B} \gets vote(B)$

                \State $\text{\aggsig} \gets \aggsig \cup \sigma_{B}$
                \If {$\textit{children} = \emptyset $}
                    \Startt {\aggtimer} \label{line:starttimer} 
                \Else\Comment{tree leaf}
                    \Send{$\langle \textsc{signature} , \sigma_{B} \rangle$}{\textit{parent}}
                    \label{line:sendsign}
                \EndIf
            \EndOn

            \vspace{0.1cm}
            
            \On $\langle \textsc{signature} , \textit{sig} \rangle$
                \label{line:onsig}
                \Assert $\text{verifies}(\text{\textit{sig}}, \textit{sig.signers})$
                \State $\text{\aggsig} \gets \aggsig \cup \textit{sig}$\label{line:aggregate}
            \EndOn

            \vspace{0.1cm}

            \On timeout(\aggtimer)
                \If{$\text{isRoot}(\textit{self})$} \label{line:leader}
                    \Comment{$root$ is $L_{v+1}$}
                    \State{$\textit{missing} \gets \Pi - \aggsig.\textit{signers}$}
                    \Send{$\langle \scndchance , B\rangle$}{\textit{missing}}
                    \Startt {\sctimer}
                    \label{line:lastchance}
                \Else \label{line:timeout!leader}
                    \Send{$\langle \textsc{signature} , \aggsig \rangle$}{\textit{parent}}\label{line:send!leader}
                \Send{$\langle \textsc{ack} , \aggsig \rangle$}{\textit{children}}\label{line:sendack}
                \EndIf
            \EndOn

            \vspace{0.1cm}

            \On $\langle \textsc{ack} , \textit{sig} \rangle$
                \Assert{$\text{verifies}(\textit{sig})$}
                \State $\aggsig \gets \textit{sig}$
            \EndOn   

            \vspace{0.1cm}
            
            \On $\langle \scndchance , B, \textit{proof}\, \rangle$ from \textit{p}
                \Assert{$\text{isValid}(B,\textit{proof}, \textit{p})$}
                \If{$B$ has new view}
                    \State $\textit{deliver}(B)$
                    \label{line:deliver2}
                    \State $\sigma_{B} \gets vote(B)$
                    \State $\text{\aggsig} \gets \aggsig \cup \sigma_{B}$
                \EndIf
                \Send{$\langle \textsc{signature} , \aggsig \rangle$}{\textit{sender}}\label{line:reply2ndch}
            \EndOn

            \vspace{0.1cm}

            \On timeout(\sctimer)\Comment{at $L_{v+1}$}
                \State $\textit{aggregate}(aggSig, \aggsig.\textit{signers})$
            \EndOn     
    \end{algorithmic}
\end{algorithm}

%% file: tex/incentives.tex
\noindent
We use game theory to analyze the possible strategies for processes in different roles. 
We model the system as a two-player game, where each player controls a fraction of the processes.
We show that if the player controlling the majority of processes acts honestly, then strategies available to the minority player are dominated by the honest strategy. 

\paragraph{Player Set}

We assume two players, an honest player $p_h$ and an attacker $p_a$.
We assume that $p_a$ controls a fraction $m< 0.5$ of all processes.

\paragraph{Strategy Set}

The strategies available to players are expressed as $S(e_l, e_v, e_a, e_p)$. 
The parameters $e_l$, $e_v$, $e_a$, and $e_p$ express different possible attacks. We omit some strategies that are obviously not beneficial. 
For example, not proposing a block since it results in zero reward.
The strategy $S_0=S(0,0,0,0)$ corresponds to correct behavior.
The attacks available to a player depend on its processes' roles in a round: round leader, internal process, and leaf process.

The leader collects signatures for the block. It can submit complete subtrees or individual, \scndchance messages from the block. Parameter $e_l$ describes a strategy in which the player tries to omit $e_l\cdot n$ many signatures belonging to the other player. To form a valid block, $e_l\leq f$ must hold.

If a player controls processes that are not the leader, these processes can refrain from voting for a block. We assume $e_v\cdot n$ many processes belonging to the player omit their votes. 

Internal processes aggregate signatures in their subtree. They may omit aggregating these signatures, leaving signatures to be aggregated by \scndchance messages instead. The player omits aggregation of $e_a\cdot n$ many signatures from processes belonging to the other player.

Leaf processes can refrain from sending their signatures to their parent, sending them in a \scndchance message to the leader instead. 
We assume $e_s\cdot n$ many processes under a player's control do this.

\paragraph{Utility Function}

We define the player's utility function as its payoff in each round. 
This payoff includes both the voting reward and the aggregation bonus. 

In the following, we analyze the profitability of different strategies for player $p_a$, assuming that $p_h$ follows $S_0$.
In any strategy $S'$ other than $S_0$, both $p_a$ looses some rewards compared to $S_0$. Let $L[S']$ be this loss. The total rewards lost by $p_a$ and $p_h$ ($R[S']$) are redistributed, and $p_a$ gains $m\cdot R[S']$.
We derive conditions, such that $m\cdot R[S'] < L[S']$, which ensures $S'$ is dominated by $S_0$.

\subsection{Vote Omission}
\noindent
A player controlling the leader may omit entire subtrees. In $S(e_l, 0, 0, 0)$ the leader omits $e_l\cdot n$ many votes, belonging to another player.
In this case, the voting reward of omitted processes $e_lb_vR$ and the aggregation reward $e_lb_aR$ for these votes
are redistributed among all processes. Similarly, the leader bonus is reduced by $\frac{e_l}{f}b_lR$ and redistributed.

With this strategy, player $p_a$ loses at least $\frac{e_l}{f}b_lR$ but gains a fraction $m$ of the redistributed rewards. We deduce the following condition:
\begin{align}
 \frac{e_l}{f}b_l &> m\left(\frac{e_l}{f}b_l+e_lb_a+e_lb_v \right)\\
 \Leftrightarrow  b_l &> \frac{mf}{1-m+mf} \label{eq:bl}
\end{align}

\subsection{Vote Denial}
\noindent
If a player is in control of non-leader processes, these may refrain from voting. In strategy $S(0,e_v,0,0)$ a player refrains from voting with $e_vn$ many of its processes.
We only consider this \textit{\novote} attack when the player does not hold the leader.
In this case, the player loses the voting reward for omitted votes $e_v\cdot b_vR$ but gains fraction $m$ of the redistributed leader bonus $\frac{e_v}{f}b_lR$ and aggregation bonus $e_vb_aR$. The lost voting reward is also redistributed.
We deduce the following condition:
\begin{align}
e_vb_v &>m(\frac{e_v}{f}b_l + e_vb_a + e_vb_v)\\
\Leftrightarrow b_l &< \frac{f(1-b_a-m)}{m+f-mf}\label{eq:novote}
\end{align}

\subsection{Aggregation Denial}
\noindent
A leaf process in the tree can not send its vote to its parent and reply to \scndchance messages instead.
We refer to this attack as \textit{\stealing}. 
We use the parameter $e_a$ for a strategy where $e_an$ many processes from the player perform this attack.
In this attack, the attacker is punished, losing $e_ab_aR$ of its voting reward. This punishment and the denied aggregation bonus $e_ab_aR$ are redistributed. 
Thus, this attack is not profitable if the following equation holds:
\begin{equation}
  m2e_ab_a < e_ab_a \label{eq:deniedagg}
\end{equation}

\subsection{Aggregation Omission}
\noindent
If a player controls an internal process, it can skip aggregating some connected leaf processes, leaving the leaf processes' votes to be collected via \scndchance messages. This will result in punishment for the leaf processes.
We refer to this attack as \textit{\forced}. 
If $e_pn$ many signatures from the leaf processes belonging to other players are not aggregated, the attacker loses $e_pb_aR$ of its aggregation reward.
The punishment and lost aggregation bonus are redistributed.
This results again in Equation~\ref{eq:deniedagg}.
For $m<0.5$, Equation~\ref{eq:deniedagg} holds and we get the following Lemma:

\begin{theorem}
For a player $p_a$ with $m<0.5$, if Equations~\ref{eq:bl} and~\ref{eq:novote} hold, then all strategies $S(e_l,e_v,e_a,e_p)$ are dominated by $S(0,0,0,0)$.
\end{theorem}

\begin{proof}
This follows from the analysis above, since the redistributed and lost rewards ($R[S']$ and $L[S']$) for different attacks sum up.
\end{proof}

%% file: tex/security.tex
\noindent
This section analyzes the security of \protocolname{} against possible attack scenarios. 

\input{./figures/simulations/simulations.tex}

\subsection{Targeted Vote Omission}
\noindent
Here we analyze the security of \protocolname{} against targeted vote omission attack with collateral $0$, in which the attacker tries to omit an individual vote.  

In \proj, the direct parent is not able to omit its children since the \scndchance messages help an omitted process to get re-added by the tree root.
Additionally, due to the indivisible multi-aggregation schemes, the root is not able to retrieve and omit one specific signature from the aggregated signatures it receives.  
Therefore, in order for the attack to be successful, the attacker needs to control two specific processes.
If the victim is a tree leaf, the attacker can omit its signature if it controls
both the root of the tree and the direct parent of its victim in one view.
Considering $m$ to denote the attacker's power as the fraction of committee members the attacker controls, and $P$ is the probability of the victim to be a leaf, the probability of such an attack is $P \cdot m^2$. 

Omitting an individual vote is also possible if the victim is an internal process, and the attacker controls both the current and previous view leaders.
In this way, the attacker can skip sharing the block proposal with the victim, and collect the victim's children through \scndchance messages.
Note that controlling both leaders is required for this scenario since the block proposal is created by the leader of the previous view and is shared with both the current view leader and its children. 
The probability of such a scenario is $(1-P) \cdot m^2$.

\begin{theorem}
\label{the:theorem1}
In \protocolname, the probability for an attacker with power $m$, to omit only its target is $m^2$.
\end{theorem}

\begin{proof}
    This is an immediate result of summing the above probabilities:
        $P \cdot m^2 + (1-P) \cdot m^2 = m^2 $
\end{proof}

\begin{corollary}
\label{lem:lemma1}
Considering the two attacks above, $0$-omission probability of \proj is $m^2$.
\end{corollary}

\noindent
Note that if an internal process does not respond with an acknowledgment to the received signatures, a process might be lured into replying to a \scndchance sent by a faulty leader and gets omitted. 
Therefore, Theorem~\ref{the:theorem1} holds if the victim receives the acknowledgment from a correct parent before a potential \scndchance from the attacker. 

An attacker can still exclude a whole branch ($a+1$ processes, considering $a$ leaves for the aggregator) to omit one targeted process by having access to $L_v$ (collateral of $a$). 
This is further analyzed in our simulations below.
However, existing incentive-based solutions are well suited to prevent such large omissions and may be applied additionally to \protocolname.

\subsection{Simulations}
\noindent

To prove the security of \protocolname{} against the mentioned attacks, we conducted different simulations.
We use Gosig and a simple star protocol with round-robin leader election as the baseline. 
Unless mentioned otherwise, in all simulations related to \proj there are $111$ processes in the committee, forming a 2-level tree with fan-out of $10$. 
Results of the simulations are shown in Figure~\ref{fig:simulations}.

We first simulated the targeted vote omission attack with collateral of $0$ in Gosig under different $k$ and different attacking power $m$. 
We also looked into situations where $30\%$ of the processes are free riding and also situations where the malicious leader tries to be greedy, and initiates the aggregation process by first sharing the signature with the victims.
As shown in Figure~\ref{fig:simulations-a}, while Gosig can defend against the attack under small $k$ and $m$, increasing these parameters highers the omission probability of Gosig that of a star protocol. 
The results also show that free riding makes the attack more successful. 
For example, while having $k = 2$ and $m = 5\%$ the attack in Gosig happens only $4\%$ of the time, free-riding increases the chances of the attack up to $24\%$.

\input{./plots/figures/evaluation_plots.tex}

In the second simulation we analyzed the robustness of \proj and Gosig against vote omission attack under different collateral. 
Figure~\ref{fig:simulations-b} shows the number of successful omissions based on the collateral. 
In this simulation, the attacking power $m$ is set to $5\%$. 
Different than Gosig, collateral has little effect on omission probability in \proj, as long as it is not enough to allow removal of a complete sub-tree.
Thus \proj has a reasonable and mostly better omission probability compared to baseline methods under different collateral. 

The third simulation compares the fraction of the reward lost by victim and attacker 
under different attacks in \proj, with the star protocol as the baseline. 
In \proj, we use $b_l$ as $15\%$, and $b_a$ as $2\%$.
The baseline also uses the same leader bonus, but not aggregation reward.

Figure~\ref{fig:simulations-c} shows the difference between the reward gained by the victim and attackers with their expected share ($1/111$). 
We can see that while in baseline, an attacker with $m = 0.3$ is able to lower the expected share of the victim by vote omission attack almost $25\%$, in \protocolname{} this is reduced to around $7\%$.
The effect of \novote attack is almost the same in both baseline and \proj, but it's is a much more expensive attack compare to vote omission, since the attackers lose much more for performing the attack.
We note that, while for a larger attacker, the fraction of reward lost in the attack is reduced, the actual cost still increases. 

In the fourth simulation we show the effect of the tree configuration (number of internal processes) on vote omission with any collateral. 
Figure~\ref{fig:simulations-d} compares how much reward (percentage of the block reward) attacker and victim lose in \proj having 4 and 10 internal processes (111 and 109 processes in total respectively), and star protocol as the baseline. 
For example, an attacker with $m = 0.1$ loses $7$ times more in \proj with 10 internal processes compared to the star protocol. 
Having larger sub-trees makes the attack with high collateral even more expensive due to the larger number of children under each aggregator. 
We see that an attacker with $m = 0.1$ loses $15$ times more in \proj with 4 internal processes compared to the baseline. 
This shows while \proj is unable to reduce the probability of the attack for higher collateral, it effectively increases the cost of the attack, making it more difficult to perform.

%% file: figures/simulations/simulations.tex
\begin{figure*}
    \centering
    \begin{subfigure}[t]{0.49\textwidth}
        \centering
        \includegraphics[width=\linewidth]{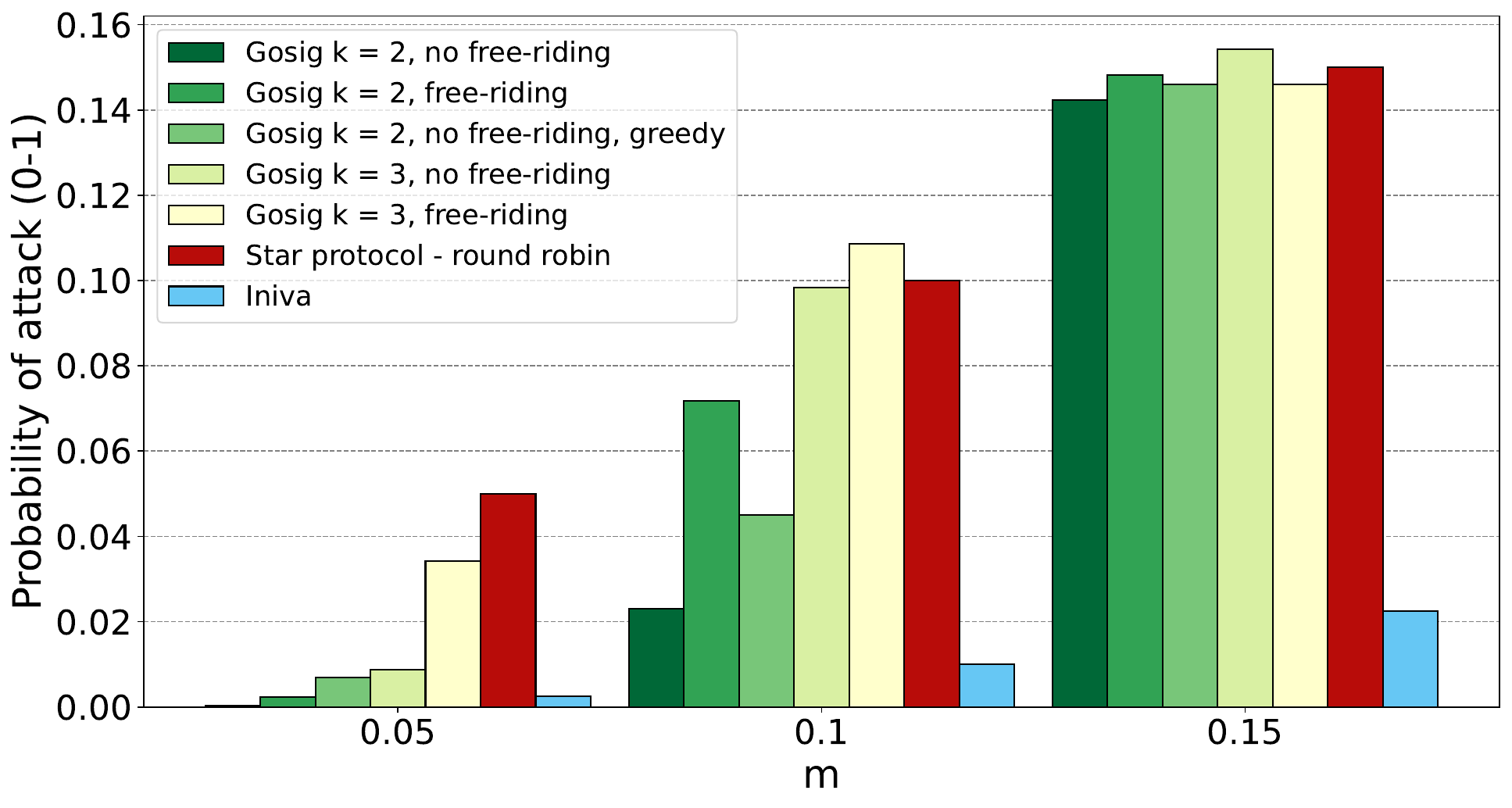}
        \caption{Vote omission probability with collateral 0: This simulation compares the possibility of targeted vote omission attack in Gosig with Star protocol and \proj under different values of $k$ and $m$.
        }
        \label{fig:simulations-a}
    \end{subfigure}
    \hfill
    \begin{subfigure}[t]{0.49\textwidth}
        \centering
        \includegraphics[width=\linewidth]{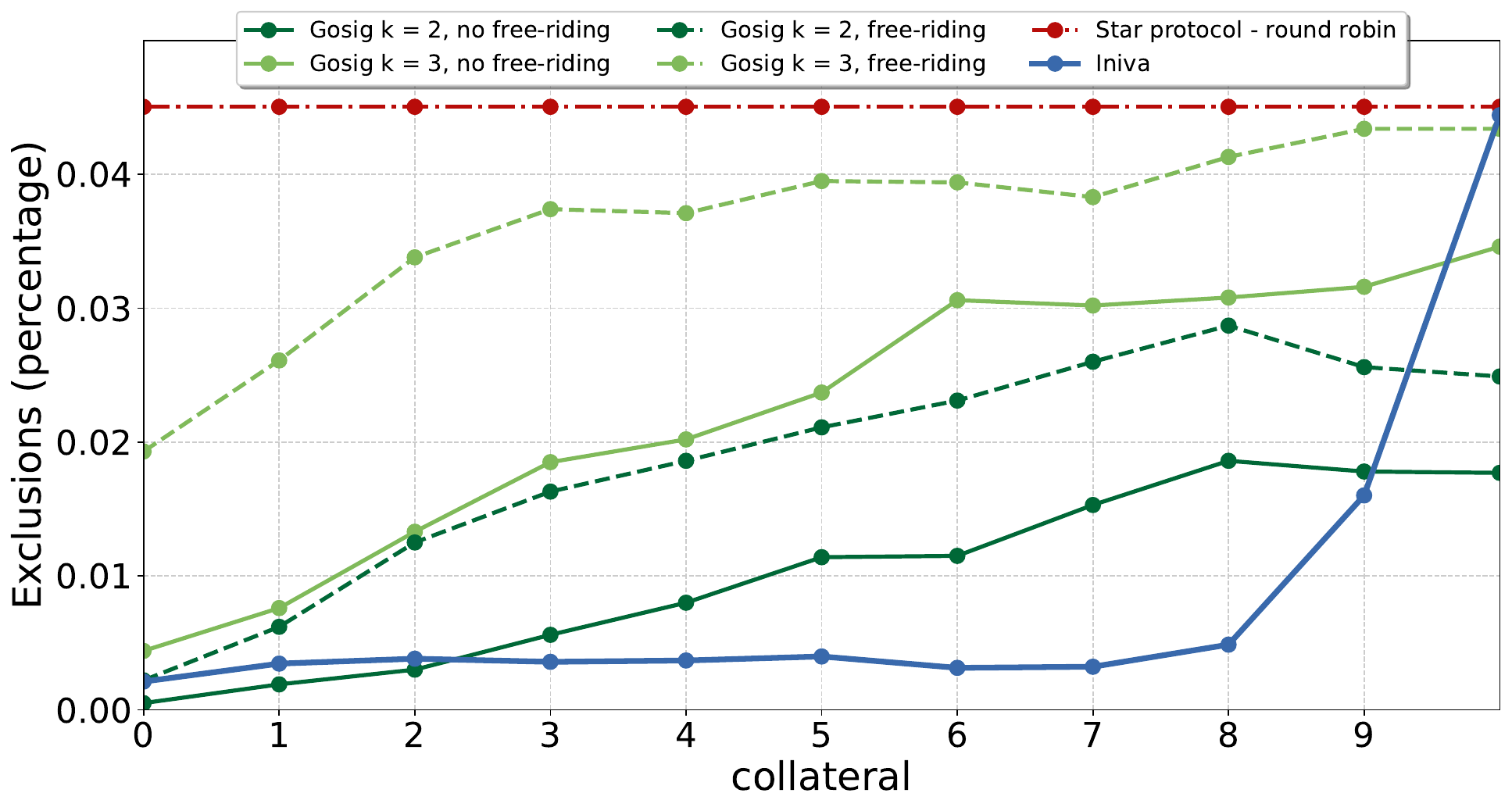}
        \caption{Vote omission probability with larger collateral: Percentage of blocks with successful vote omission attack in \proj, Gosig, and Star protocol with different collateral. 
        $m$ is set to $5\%$.}
        \label{fig:simulations-b}
    \end{subfigure}
    \hfill
    \begin{subfigure}[t]{0.58\textwidth}
        \centering
        \includegraphics[width=\linewidth]{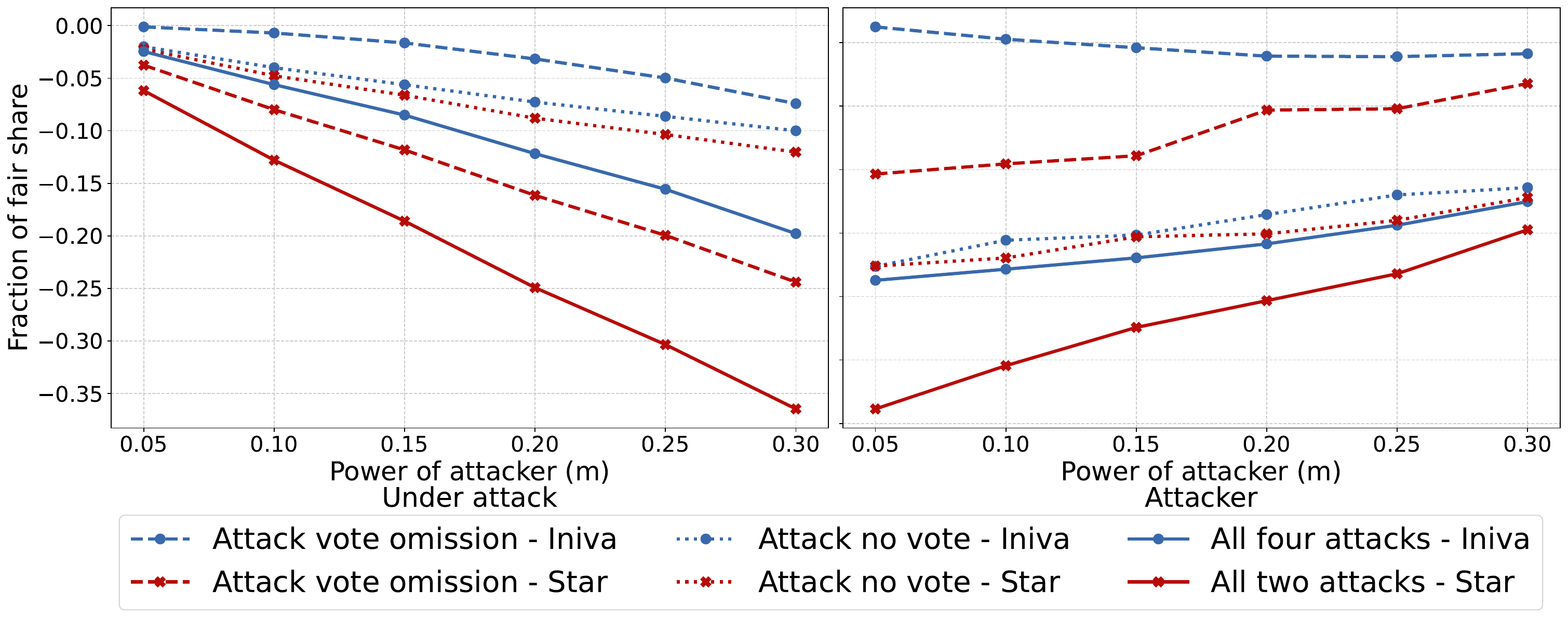}
        \caption{Effect of different attacks with collateral 0: It compares the fraction of the fair reward lost by the victim and attacker under different attacks under different ($m$) for a star protocol with leader reward and \proj.
        }
        \label{fig:simulations-c}
    \end{subfigure}
    \hfill
    \begin{subfigure}[t]{0.40\textwidth}
        \centering
        \includegraphics[width=\linewidth]{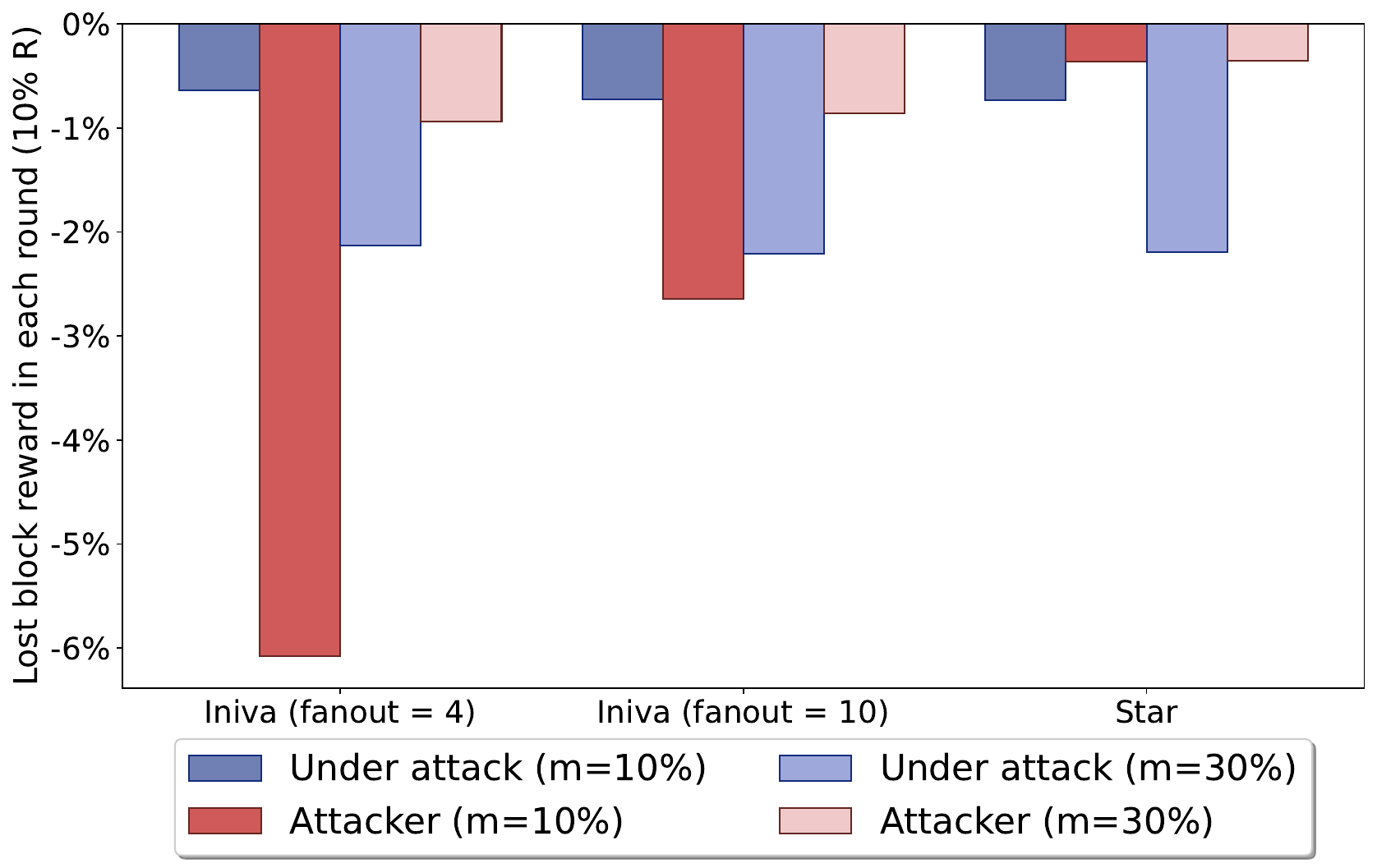}
        \caption{Vote omission effect with large collateral: Compares the reward lost (a percentage of the block reward) for both the attacker and the victim when the attacker removes up to a whole branch for omitting its target in \proj with 4 and 10 internal nodes and the star protocol. }
        \label{fig:simulations-d}
    \end{subfigure}
    \hfill
    \caption{Simulation results. In each simulation, there are $111$ processes in each committee for \proj (a full 2-level tree with a fan-out of 10). In (a) and (b) there are $100$ processes in the committee for Gosig. In (d), there are $109$ processes when having 4 internal nodes.
    }
    \label{fig:simulations}
\end{figure*}

%% file: plots/figures/evaluation_plots.tex
\begin{figure*}[h]

    \centering
    \begin{subfigure}[b]{0.32\textwidth}
        \centering
        \resizebox{\linewidth}{!}{%
        \input{plots/baseval2.tex}
        }
       \caption{Throughput vs latency for HotStuff, \proj, and \proj-No2C.}
        \label{fig:baseperf}
    \end{subfigure}
    \begin{subfigure}[b]{0.33\textwidth}
    \centering
        \resizebox{\linewidth}{!}{%
        \input{plots/cpubls12.tex}
        }
       \caption{CPU usage of HotStuff and \proj.}
    \label{fig:bls12cpu}
    \end{subfigure}
    \begin{subfigure}[b]{0.32\textwidth}
        \centering
        \resizebox{\linewidth}{!}{%
        \input{plots/config_th.tex}
        }
       \caption{Scalability evaluation with batch size $B=100$ and
        varying configuration sizes and payloads.
        Configuration sizes are selected to have an almost complete tree of height 3.}
        \label{fig:configlth}
    \end{subfigure}
    \caption{Experimental results with 21 replicas, 4 clients and different payload and batch sizes.}
    \label{fig:evaluation_plots}
\end{figure*}
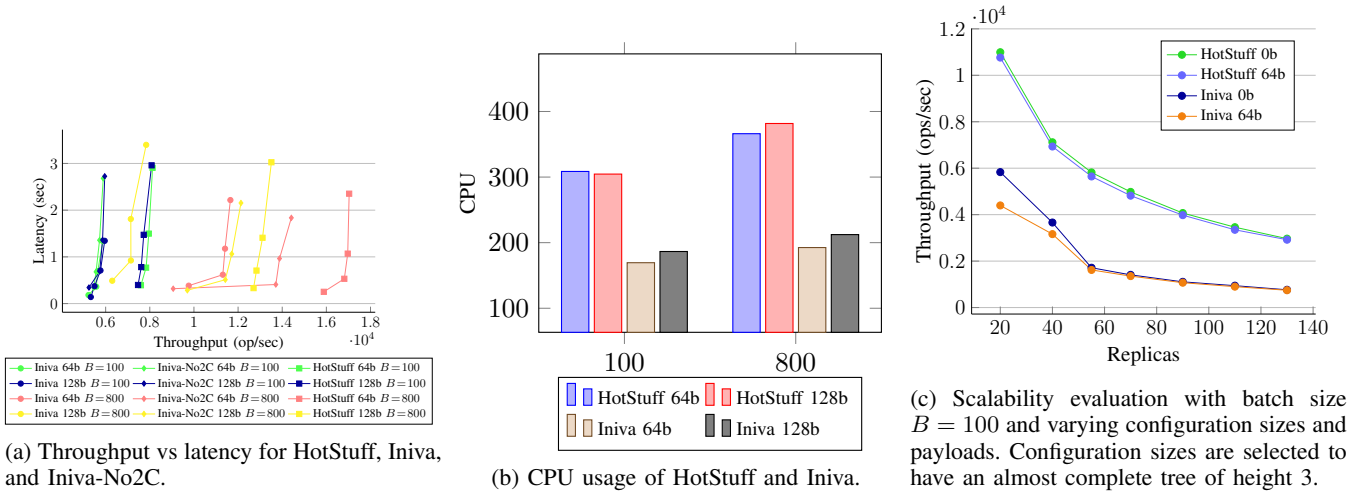

%% file: plots/baseval2.tex
\begin{tikzpicture}
\def\datafileA{plots/new_data/randelnf-final-100-64.csv}
\pgfplotstableread{\datafileA}\datatable
\def\datafileB{plots/new_data/randelnf-final-100-128.csv}
\pgfplotstableread{\datafileB}\datatable
\def\datafileC{plots/new_data/hotstuff-final-100-64-sync.csv}
\pgfplotstableread{\datafileC}\datatable
\def\datafileD{plots/new_data/hotstuff-final-100-128-sync.csv}
\pgfplotstableread{\datafileD}\datatable

\def\datafileE{plots/new_data/randelnf-final-800-64.csv}
\pgfplotstableread{\datafileE}\datatable
\def\datafileF{plots/new_data/randelnf-final-800-128.csv}
\pgfplotstableread{\datafileF}\datatable
\def\datafileG{plots/new_data/hotstuff-final-800-64-sync.csv}
\pgfplotstableread{\datafileG}\datatable
\def\datafileH{plots/new_data/hotstuff-final-800-128-sync.csv}
\pgfplotstableread{\datafileH}\datatable

\def\datafileI{plots/new_data/randel-final-100-64.csv}
\pgfplotstableread{\datafileI}\datatable
\def\datafileJ{plots/new_data/randel-final-100-128.csv}
\pgfplotstableread{\datafileJ}\datatable
\def\datafileK{plots/new_data/randel-final-800-64.csv}
\pgfplotstableread{\datafileK}\datatable
\def\datafileL{plots/new_data/randel-final-800-128.csv}
\pgfplotstableread{\datafileL}\datatable
\begin{axis}[
    tick label style={font=\large},
    label style={font=\large},
    legend style={at={(0.5,-0.25)},
	legend cell align={left},
	anchor=north,legend columns=3,font=\small},
	axis lines*=left,
	ymajorgrids,
	width=10cm, height=6.5cm,
	xlabel={Throughput (op/sec)},
	ylabel={Latency (sec)},
 y label style={yshift=-18pt}, 
	legend entries={
\proj 64b $B\!=\!100$,
\proj-No2C 64b $B\!=\!100$,
HotStuff 64b $B\!=\!100$,
\proj 128b $B\!=\!100$,
\proj-No2C 128b $B\!=\!100$,
HotStuff 128b $B\!=\!100$,
\proj 64b $B\!=\!800$,
\proj-No2C 64b $B\!=\!800$,
HotStuff 64b $B\!=\!800$,
\proj 128b $B\!=\!800$,
\proj-No2C 128b $B\!=\!800$,
HotStuff 128b $B\!=\!800$,  },
 	cycle multi list={%
{green!70!white, solid, mark=*},
{green!70!white, solid, mark=diamond*},
{green!70!white, solid, mark=square*},
{blue!60!black, solid, mark=*},
{blue!60!black, solid, mark=diamond*},
{blue!60!black, solid, mark=square*},
{red!50!white, solid, mark=*},
{red!50!white, solid, mark=diamond*},
{red!50!white, solid, mark=square*},
{yellow!90!white, solid, mark=*},
{yellow!90!white, solid, mark=diamond*},
{yellow!90!white, solid, mark=square*},}
	]
 \addplot+ table[x=throughput, y=latency, col sep=comma]{\datafileI};
	\addplot+ table[x=throughput, y=latency, col sep=comma]{\datafileA};
 \addplot+ table[x=throughput, y=latency, col sep=comma]{\datafileC};

  \addplot+ table[x=throughput, y=latency, col sep=comma]{\datafileJ};
	\addplot+ table[x=throughput, y=latency, col sep=comma]{\datafileB};
 \addplot+ table[x=throughput, y=latency, col sep=comma]{\datafileD};

 \addplot+ table[x=throughput, y=latency, col sep=comma]{\datafileK};
\addplot+ table[x=throughput, y=latency, col sep=comma]{\datafileE};
\addplot+ table[x=throughput, y=latency, col sep=comma]{\datafileG};

\addplot+ table[x=throughput, y=latency, col sep=comma]{\datafileL};
\addplot+ table[x=throughput, y=latency, col sep=comma]{\datafileF};
\addplot+ table[x=throughput, y=latency, col sep=comma]{\datafileH};

\end{axis}
\end{tikzpicture}

%% file: plots/cpubls12.tex
\begin{tikzpicture}
\begin{axis}[
	label style={font=\footnotesize},
    legend cell align={left},
    x tick label style={/pgf/number format/1000 sep=},
    ylabel=CPU,
    xlabel={Batch size},
    width=\linewidth,
    enlargelimits=0.5,
    legend style={at={(0.5,-0.16)}, anchor=north,legend columns=2, font=\scriptsize},
    ybar,
    xtick=data,
    y label style={yshift=-9pt} 
]
\addplot coordinates {(100,308.5705195759) (800,366.133502049248)};
\addplot coordinates {(100,304.54874803872) (800, 381.6882764672)};
\addplot coordinates {(100,169.43675303538078) (800,192.54979880755332)};
\addplot coordinates {(100,186.52366990596605) (800,212.30082951625786)};
\legend{HotStuff 64b, HotStuff 128b, Iniva 64b, Iniva 128b}
\end{axis}
\end{tikzpicture}

%% file: plots/config_th.tex

\begin{tikzpicture}
\def\datafileA{plots/new_data/hotstuff-no-payload-th.csv}
\pgfplotstableread{\datafileA}\datatable
\def\datafileB{plots/new_data/hotstuff-payload-th.csv}
\pgfplotstableread{\datafileB}\datatable
\def\datafileC{plots/data/config/randel-no-payload-th.csv}
\pgfplotstableread{\datafileC}\datatable
\def\datafileD{plots/data/config/randel-payload-th.csv}
\pgfplotstableread{\datafileD}\datatable

\begin{axis}[
    tick label style={font=\large},
    label style={font=\large},
	legend cell align={left},
    legend style={at={(0.55,.80)},anchor=west,font=\small},
	axis lines*=left,
	ymajorgrids,
	xlabel={Replicas},
	ylabel={Throughput (ops/sec)},
    y label style={yshift=-9pt}, 
	legend entries={HotStuff 0b, HotStuff 64b, Iniva 0b,Iniva 64b},
 	cycle multi list={
		{solid, mark=*},{densely dashed, mark=square*},{solid, mark=triangle*}, {densely dashed, mark=diamond*}\nextlist
		green!70!gray,blue!60!white,blue!60!black,
		orange!90!gray,yellow!90!gray,red!90!gray}
	]
	\addplot+ table[x=nodes, y=throughput, col sep=comma]{\datafileA};
	\addplot+ table[x=nodes, y=throughput, col sep=comma]{\datafileB};
	\addplot+ table[x=nodes, y=throughput, col sep=comma]{\datafileC};
	\addplot+ table[x=nodes, y=throughput, col sep=comma]{\datafileD};
\end{axis}
\end{tikzpicture}

%% file: tex/experiments.tex
\subsection{Implementation}
\noindent
We implemented \proj, integrating the signature aggregation described in Algorithm~\ref{alg:sig} in an existing implementation of the HotStuff consensus algorithm~\cite{appliedHotstuff}~\footnote{The source code for the experiments and simulations is available at \url{https://github.com/relab/iniva-artifacts}.}.
\proj is added as a module in the framework to perform propagation of blocks and vote aggregation.
\proj does not change the implementation of consensus, or client and request handling.

The HotStuff algorithm operates in synchronous rounds~\cite{yin2019hotstuff}.
A new block is only proposed after the votes for the previous block have been aggregated.
In this setting, additional latency during dissemination and waiting for additional votes affects not only latency but also the throughput of the protocol.
This allows us to realistically evaluate the overhead added by \proj.

We also implemented a few variants of \proj to evaluate our design choices.
In most BFT protocols, the leader stops collecting/waiting for votes once it has a quorum.
\proj triggers a \scndchance after obtaining a QC to provide a second chance to the processes which their parents intentionally left out.
To understand the overhead of this design choice, we implemented a variant that we call \proj-No2C, where no \scndchance messages are sent.
\proj-No2C provides the cost of proposal dissemination and vote aggregation in the tree communication model.

\input{./plots/figures/faults_plot.tex}

The \aggtimerl started on Line~\ref{line:starttimer} of Algorithm~\ref{alg:sig} determines performance and inclusion of the protocol.
If the timer is set too low, the leader may not be able to collect a QC, causing a view failure.
If it is set too high, the processes will wait longer for the contribution from faulty processes, resulting in degraded performance.
For failure scenarios, we varied the timer to understand its effect on view failures, throughput, latency, and inclusion.

\subsection{Setup}
\noindent
We used our local cluster to evaluate our implementation.
The cluster contains 25~machines and each node has 32~GB of RAM and 12~cores of Intel Xenon processors with a maximum frequency of 3.3~GHz.
A 10~Gbps TOR switch connects nodes and the latency among the nodes is less than 1~ms.
We used round-robin leader rotation policy in the experiments, except for a few experiments where we used the Carousel leader election policy~\cite{cohen2022carousel}.
All experiments run for 150 seconds and metrics are collected every second.
The first 5 seconds are used as a warm-up period.
All results have less than 1\% variance with a 90\% confidence interval.

\subsection{Evaluation}
\noindent
We evaluated our implementation in three ways, each with a different objective.
\begin{itemize}
    \item Base evaluation is performed to evaluate the overhead, throughput, latency, and resource utilization of \proj and compare it with HotStuff in a fault-free configuration.
    \item Scaling experiments are conducted to compare \proj and HotStuff with increasing configuration size.
    \item Resiliency evaluation is conducted on the different \proj variants to understand the effect of failures on throughput, latency, and inclusiveness.
\end{itemize}

\subsubsection{Base Evaluation}
We used 21~machines as processes and 4~machines as clients.
For \proj, these 21~processes are arranged as a complete tree of height~2 with 4~internal and 16~leaf nodes.

Clients send the request to all processes and expect a quorum of replies before considering the request committed.
Requests contain 64 or 128 bytes payload.
Batching of requests is enabled at the processes and we used 100 and 800 batch sizes for this evaluation.
Clients measure latency and the throughput is measured at the processes.
We used BLS12~\cite{bls12} for signature aggregation.

Figure~\ref{fig:baseperf} shows throughput and latency under different client loads.
The \aggtimerl is adjusted based on the client load on the cluster.
We observe that the throughput of \proj is $\sim33\%$ lower than HotStuff.
The tree-based communication without \scndchance (\proj-No2C) is responsible for about half of the overhead.
Although throughput is not the primary objective of \proj, it can be compensated for with larger batch sizes.
Additionally, pipelining of requests in the tree, similar to Kauri~\cite{neiheiser2021kauri} could improve throughput.
Also, \proj still has a reasonable throughput compared to most \ac{PoW}-based schemes such as Bitcoin.

Figure~\ref{fig:bls12cpu} shows the CPU usage for HotStuff and \proj for two different payload sizes (64 and 128 bytes) and batch sizes ($B=100$ and $B=800$).
The CPU usage is measured as the percentage of CPU time used by the process.
The results show that \proj uses $\sim48\%$ less CPU compared to HotStuff.
The lower CPU consumption is due to \proj's tree structure.
The tree structure distributes the load and thus reduces CPU usage, but also increases latency and reduces throughput.
Doubling the payload from 64 to 128 bytes does not significantly impact CPU usage.
When the throughput results are correlated with the CPU usage, we argue that \proj could outperform HotStuff in a resource-constrained environment.

\subsubsection{Scaling Evaluation}

To evaluate the scalability, we run up to 130 processes, having each physical machine hosting 5~processes.
We use batch size 100 and 4 clients.
With increased configuration size, the branching factor of the tree is increased to keep the tree's height constant.
Figure~\ref{fig:configlth} shows throughput observed for various configurations with and without payload for HotStuff and \proj.
With increased configuration size, throughput decreases gradually.

\subsubsection{Resiliency Evaluation}

We conducted the resiliency evaluation of the \proj protocol by inducing crash failures in the configuration.
As explained earlier, \proj reconfigures the position of the processes in the tree for every view and faulty processes are randomly placed in the tree.
The experiment is done with 21 processes, each running on individual machines with batch size 100 and 4 clients.
We set the \aggtimerl and \sctimerl based on the following heuristic.
Let $\Delta$ be the network delay between the processes.
The \aggtimerl is set to $2\Delta\cdot \text{height}(p)$, where $\text{height}(p)$ is $p$'s height in the tree.
The \sctimerl is set to $\delta=2\Delta$.
We repeated the experiments with two different $\delta$ values, 5ms and 10ms.

Figure~\ref{fig:faults_experiments} shows the effect of failures on the throughput, latency, failed views, and inclusion.
With faulty processes in the system, internal processes will wait for votes and the leader will wait for \scndchance messages.
With increasing failures, latency increases and throughput decreases, as seen in Figures~\ref{fig:faultsthroughput} and \ref{fig:faultslatency}.
The longer \sctimerl of 10~ms causes higher latencies and lower throughput.

Figure~\ref{fig:failedviews} shows the percentage of failed views. A view may fail either because its leader is faulty, or because no QC could be collected.
We also included a variant of \proj that uses the Carousel leader election to avoid electing faulty leaders.
If two of the four internal processes are faulty, no QC can be collected without the \scndchance messages.
With a higher \sctimerl, the number of failed views decreased by 10\%.

One of the main objectives of the \proj  mechanism is inclusion.
Figure~\ref{fig:inclusion} shows the average number of votes included.
With 4~failures \proj includes more than 99\% of correct processes.
Our baseline, HotStuff, always includes a quorum of 15 votes.
We also see that the increased timer has a positive effect on inclusion.

%% file: plots/figures/faults_plot.tex
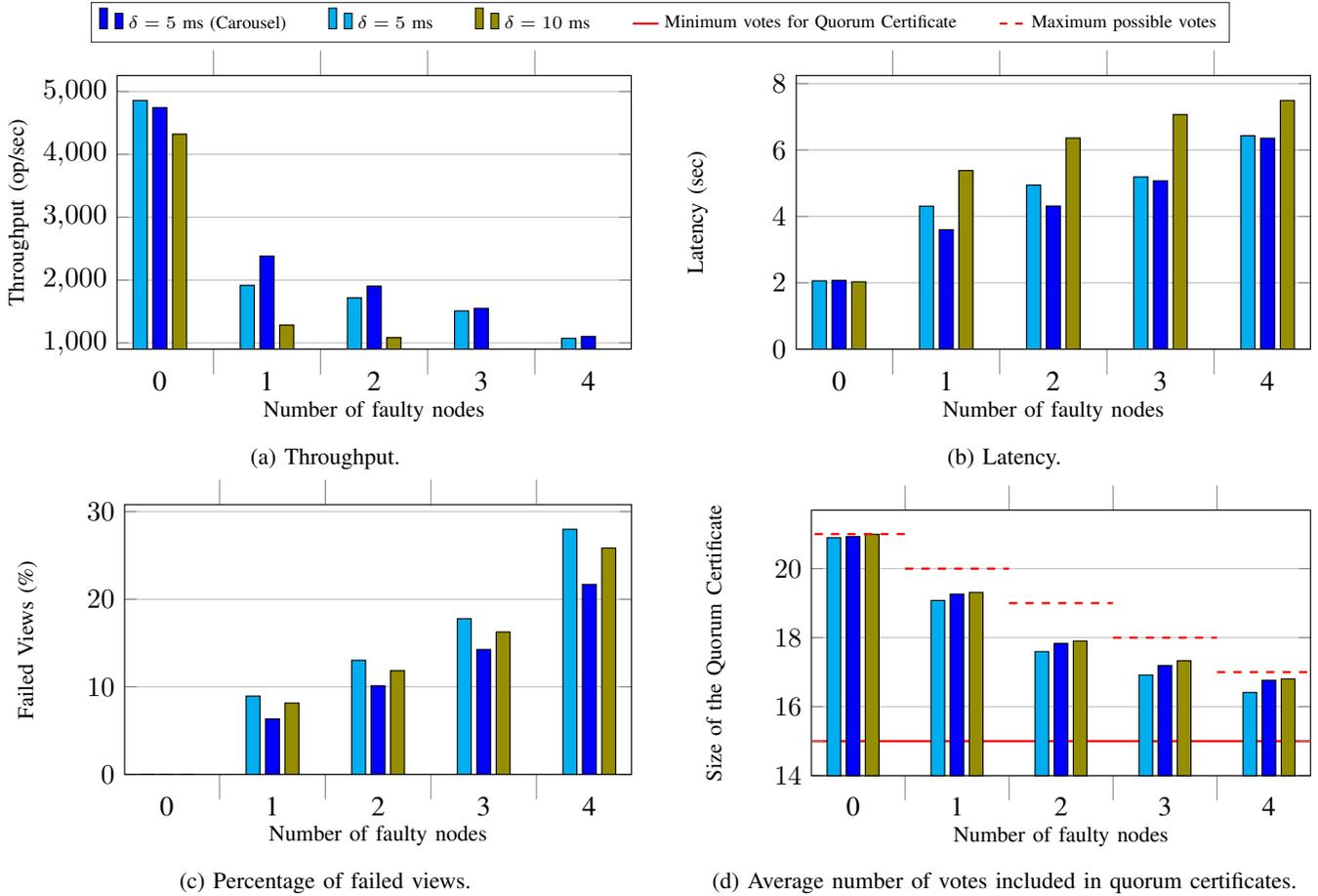
\begin{figure*}[h]

    \centering

    \begin{subfigure}[b]{0.89\textwidth}
        \centering
        \input{plots/figures/legend_faults.tex}
    \end{subfigure}\hfill

    \vspace{0.1cm}
    
    \begin{subfigure}[b]{0.49\textwidth}
        \centering
        \resizebox{\linewidth}{!}{%
        \input{plots/newfaultsth.tex}
        }
        \caption{Throughput.} 
        \label{fig:faultsthroughput}
    \end{subfigure} \hfill
    \begin{subfigure}[b]{0.49\textwidth}
        \centering
        \resizebox{\linewidth}{!}{%
        \input{plots/newfaultslt.tex}
        }
        \caption{Latency.} 
        \label{fig:faultslatency}
    \end{subfigure}

    \vspace{0.1cm}
    
     \begin{subfigure}[b]{0.49\textwidth}
        \centering
        \resizebox{\linewidth}{!}{
        \input{plots/newfaultsfv.tex}
        }
        \caption{Percentage of failed views.} 
        \label{fig:failedviews}
    \end{subfigure} \hfill
    \begin{subfigure}[b]{0.49\textwidth}
        \centering
        \resizebox{\linewidth}{!}{
        \input{plots/newfaultsip.tex}
        }
        \caption{Average number of votes included in quorum certificates.}
        \label{fig:inclusion}
    \end{subfigure}
    
    \caption{Experiments with a 21-replica configuration with faulty nodes randomly placed in the tree. We vary the \sctimerl ($\delta$) and leader election policy (Round-Robin and Carousel).}
    \label{fig:faults_experiments}
\end{figure*}

%% file: plots/figures/legend_faults.tex
\pgfplotsset{
/pgfplots/my legend/.style={
legend image code/.code={
\draw[thick,red](-0.05cm,0cm) -- (0.3cm,0cm);%
   }
  }
}

\pgfplotsset{
/pgfplots/my legend2/.style={
legend image code/.code={
\draw[thick,red,dashed](-0.05cm,0cm) -- (0.3cm,0cm);%
   }
  }
}

\begin{tikzpicture} 
    \begin{axis}[%
    xmin=10,
    xmax=\textwidth,
    ymin=0,
    ymax=0.6,
    legend style={draw=white!15!black,legend cell align=center},
    legend columns=5,
    legend style={/tikz/every even column/.append style={column sep=0.5cm}},
    legend style={at={(0,0)},anchor=center,font=\scriptsize},
    scale only axis,width=1mm,
    hide axis,
    ]
    \addlegendimage{black,fill=blue,ybar,ybar legend}
    \addlegendentry{$\delta=5$~ms (Carousel)};
    \addlegendimage{black,fill=cyan,ybar,ybar legend}
    \addlegendentry{$\delta=5$~ms};
    \addlegendimage{black,fill=olive,ybar,ybar legend}
    \addlegendentry{$\delta=10$~ms};
    \addlegendimage{my legend}
    \addlegendentry{Minimum votes for Quorum Certificate}
    \addlegendimage{my legend2}
    \addlegendentry{Maximum possible votes}
    \end{axis}
\end{tikzpicture}

%% file: plots/newfaultsth.tex
\begin{tikzpicture}
    \begin{axis}[
        width  = 8cm,
        height = 5cm,
        major x tick style = transparent,
        ybar,
        bar width=5pt,
        ymajorgrids = true,
        ylabel = {Throughput (op/sec)},
        symbolic x coords={0,1,2,3,4},
        xtick = data,
        xlabel = {Number of faulty nodes},
        major x tick style = {opacity=0},
        minor x tick num = 1,
        minor tick length=2ex,
        every node near coord/.append style={
                anchor=west,
                rotate=90
        },
        ymin=900,
        label style={font=\footnotesize},
    ]
        \addplot[style={fill=cyan,mark=none}]
            coordinates {(0,4857.471160912916)(1, 1914.936218763881) (2, 1716.645768474721
            )
            (3, 1508.72174917849) (4, 1072.8375804055374)};
        \addplot[style={fill=blue,mark=none}]
             coordinates {(0,4743.092569273653) (1, 2382.308697716243)
             (2, 1904.361238578588 ) (3, 1550.125968757317) (4, 1102.9571025471932)};
        \addplot[style={fill=olive,mark=none}]
             coordinates {(0,4321.567305217373) (1, 1282.8961951625827)
             (2, 1084.1614172969191) (3, 868.083694182994) (4, 651.6534070795018)};
    \end{axis}
\end{tikzpicture}

%% file: plots/newfaultslt.tex
\begin{tikzpicture}
    \begin{axis}[
        width  = 8cm,
        height = 5cm,
        major x tick style = transparent,
        ybar,
        bar width=5pt,
        ymajorgrids = true,
        ylabel = {Latency (sec)},
        symbolic x coords={0,1,2,3,4},
        xtick = data,
        xlabel = {Number of faulty nodes},
        major x tick style = {opacity=0},
        minor x tick num = 1,
        minor tick length=2ex,
        every node near coord/.append style={
                anchor=west,
                rotate=90
        },
        ymin=0,
        label style={font=\footnotesize},
    ]
        \addplot[style={fill=cyan,mark=none}]
            coordinates {(0,2.0586481605792915)(1,  4.307892769031754) (2, 4.94250594564516)
            (3, 5.188311514493208) (4, 6.430393372468007)};
        \addplot[style={fill=blue,mark=none}]
             coordinates {(0,2.0734036618667577) (1, 3.599547274096314)
             (2, 4.311994624687306) (3, 5.0706558809153748) (4, 6.355776521110782)};
        \addplot[style={fill=olive,mark=none}]
             coordinates {(0, 2.0291173083026446) (1, 5.382277379345183)
             (2, 6.360803567788486) (3, 7.065553962482015)
             (4,  7.492388416695755)};
    \end{axis}
\end{tikzpicture}

%% file: plots/newfaultsfv.tex
\begin{tikzpicture}
    \begin{axis}[
        width  = 8cm,
        height = 5cm,
        major x tick style = transparent,
        ybar,
        bar width=5pt,
        ymajorgrids = true,
        ylabel = {Failed Views (\%)},
        symbolic x coords={0,1,2,3,4},
        xtick = data,
        xlabel = {Number of faulty nodes},
        major x tick style = {opacity=0},
        minor x tick num = 1,
        minor tick length=2ex,
        every node near coord/.append style={
                anchor=west,
                rotate=90
        },
        ymin=0,
        label style={font=\footnotesize},
    ]
        \addplot[style={fill=cyan,mark=none}]
            coordinates {(0,0.0003641408587698)(1,  8.94042678707179) (2, 13.03203019769197)
            (3, 17.7805079278675) (4, 27.98884433884206)};
        \addplot[style={fill=blue,mark=none}]
             coordinates {(0, 0.0003641408587698) (1,6.33020606314474)
             (2, 10.12616391128397) (3, 14.26045736273095) (4, 21.68920006056546)};
        \addplot[style={fill=olive,mark=none}]
             coordinates {(0,0.0002516129716322) (1, 8.14859419030432)
             (2, 11.84128041114716) (3, 16.25748256870765)
             (4, 25.84949857466783 )};
    \end{axis}
\end{tikzpicture}

%% file: plots/newfaultsip.tex
\pgfplotsset{
/pgfplots/my legend/.style={
legend image code/.code={
\draw[thick,red](-0.05cm,0cm) -- (0.3cm,0cm);%
   }
  }
}
\begin{tikzpicture}
    \begin{axis}[
        width  = 8cm,
        height = 5cm,
        major x tick style = transparent,
        ybar,
        bar width=5pt,
        ymajorgrids = true,
        ylabel = {Size of the Quorum Certificate},
        symbolic x coords={0,1,2,3,4},
        xtick = data,
        xlabel = {Number of faulty nodes},
        major x tick style = {opacity=0},
        minor x tick num = 1,
        minor tick length=2ex,
        every node near coord/.append style={
                anchor=west,
                rotate=90
        },
        ymin=14,
        label style={font=\footnotesize},
        extra y ticks = 15,
    extra y tick labels={},
    extra y tick style={grid=major,major grid style={thick,draw=red}},
    ]
        \addplot[style={fill=cyan,mark=none}]
            coordinates {(0,20.89427127996112)(1,  19.077018633540372) (2, 17.59471890971039)
            (3, 16.916507177033492) (4,  16.411815561959654)};
        \addplot[style={fill=blue,mark=none}]
             coordinates {(0, 20.926844988720593) (1, 19.260782347041122)
             (2, 17.831117021276597) (3, 17.19047619047619) (4, 16.761603375527425)};
        \addplot[style={fill=olive,mark=none}]
             coordinates {(0, 20.995277449822904) (1, 19.311627906976746)
             (2, 17.90308370044053) (3, 17.32972972972973)
             (4, 16.804214559386972)};
        \addplot[red,dashed,thick,sharp plot,update limits=false,] coordinates { ([normalized]-1,21) ([normalized]0.5,21) };
        \addplot[red,dashed,thick,sharp plot,update limits=false,] coordinates { ([normalized]0.5,20) ([normalized]1.5,20) };
        \addplot[red,dashed,thick,sharp plot,update limits=false,] coordinates { ([normalized]1.5,19) ([normalized]2.5,19) };
        \addplot[red,dashed,thick,sharp plot,update limits=false,] coordinates { ([normalized]2.5,18) ([normalized]3.5,18) };
        \addplot[red,dashed,thick,sharp plot,update limits=false,] coordinates { ([normalized]3.5,17) ([normalized]4.5,17) };
        
    \end{axis}
\end{tikzpicture}

%% file: tex/conclusion.tex
\noindent
In this paper we proposed \proj, a vote aggregation protocol to defend against targeted vote omission attacks. 
\proj is built upon \textit{Indivisibility}, a feature of some multi-signature aggregation schemes that we defined. 
Using a tree overlay and fallback paths, \proj stays inclusive and fault-tolerant. 
The designed rewarding mechanism motivates processes to participate in the aggregation procedure, and makes \proj incentive compatible. 
We conducted several experiments and simulations to analyze \proj from different perspectives such as security, throughput, latency, recourse efficiency, scalability, and tolerating faults. 
The results show while \proj outperforms previous work in terms of preventing vote omission attacks, it has a reasonable performance even in presence of faulty processes in the system.

%% file: main.bbl
\begin{thebibliography}{10}
\providecommand{\url}[1]{#1}
\csname url@samestyle\endcsname
\providecommand{\newblock}{\relax}
\providecommand{\bibinfo}[2]{#2}
\providecommand{\BIBentrySTDinterwordspacing}{\spaceskip=0pt\relax}
\providecommand{\BIBentryALTinterwordstretchfactor}{4}
\providecommand{\BIBentryALTinterwordspacing}{\spaceskip=\fontdimen2\font plus
\BIBentryALTinterwordstretchfactor\fontdimen3\font minus \fontdimen4\font\relax}
\providecommand{\BIBforeignlanguage}[2]{{%
\expandafter\ifx\csname l@#1\endcsname\relax
\typeout{** WARNING: IEEEtran.bst: No hyphenation pattern has been}%
\typeout{** loaded for the language `#1'. Using the pattern for}%
\typeout{** the default language instead.}%
\else
\language=\csname l@#1\endcsname
\fi
#2}}
\providecommand{\BIBdecl}{\relax}
\BIBdecl

\bibitem{buterin2018ethereum}
V.~Buterin, ``{E}thereum 2.0 spec--{C}asper and sharding,'' \emph{Available [online].[Accessed: 30-10-2018]}, 2018.

\bibitem{themerge}
\BIBentryALTinterwordspacing
Ethereum, ``The merge,'' Nov 2022. [Online]. Available: \url{https://ethereum.org/en/upgrades/merge/}
\BIBentrySTDinterwordspacing

\bibitem{kwon2016cosmos}
J.~Kwon and E.~Buchman, ``Cosmos: A network of distributed ledgers,'' \emph{URL https://cosmos. network/whitepaper}, 2016.

\bibitem{chen2019algorand}
J.~Chen and S.~Micali, ``Algorand: A secure and efficient distributed ledger,'' \emph{Theoretical Computer Science}, vol. 777, pp. 155--183, 2019.

\bibitem{liu2020fairselection}
Y.~Liu, J.~Liu, Z.~Zhang, and H.~Yu, ``A fair selection protocol for committee-based permissionless blockchains,'' \emph{Computers \& Security}, vol.~91, p. 101718, 2020.

\bibitem{amoussou2019fairnessCommittee}
Y.~Amoussou-Guenou, A.~Del~Pozzo, M.~Potop-Butucaru, and S.~Tucci-Piergiovanni, ``On fairness in committee-based blockchains,'' \emph{arXiv preprint arXiv:1910.09786}, 2019.

\bibitem{buterin2020incentives}
V.~Buterin, D.~Reijsbergen, S.~Leonardos, and G.~Piliouras, ``Incentives in {E}thereum's hybrid {C}asper protocol,'' \emph{International Journal of Network Management}, vol.~30, no.~5, p. e2098, 2020.

\bibitem{buterin2018discouragement}
V.~Buterin, ``Discouragement attacks,'' \emph{ETH research}, 2018.

\bibitem{baloochestani2022rebop}
A.~Baloochestani, L.~Jehl, and H.~Meling, ``Rebop: Reputation-based incentives in committee-based blockchains,'' in \emph{IFIP International Conference on Distributed Applications and Interoperable Systems}.\hskip 1em plus 0.5em minus 0.4em\relax Springer, 2022, pp. 37--54.

\bibitem{cohen2022carousel}
S.~Cohen, R.~Gelashvili, L.~K. Kogias, Z.~Li, D.~Malkhi, A.~Sonnino, and A.~Spiegelman, ``Be aware of your leaders,'' in \emph{International Conference on Financial Cryptography and Data Security}.\hskip 1em plus 0.5em minus 0.4em\relax Springer, 2022, pp. 279--295.

\bibitem{bls12}
\BIBentryALTinterwordspacing
P.~S. L.~M. Barreto, B.~Lynn, and M.~Scott, ``Constructing elliptic curves with prescribed embedding degrees,'' Cryptology ePrint Archive, Paper 2002/088, 2002, \url{https://eprint.iacr.org/2002/088}. [Online]. Available: \url{https://eprint.iacr.org/2002/088}
\BIBentrySTDinterwordspacing

\bibitem{neiheiser2021kauri}
R.~Neiheiser, M.~Matos, and L.~Rodrigues, ``Kauri: Scalable {BFT} consensus with pipelined tree-based dissemination and aggregation,'' in \emph{Proceedings of the ACM SIGOPS 28th Symposium on Operating Systems Principles}, 2021, pp. 35--48.

\bibitem{kogias2016byzcoin}
E.~K. Kogias, P.~Jovanovic, N.~Gailly, I.~Khoffi, L.~Gasser, and B.~Ford, ``Enhancing {Bitcoin} security and performance with strong consistency via collective signing,'' in \emph{25th usenix security symposium (usenix security 16)}, 2016, pp. 279--296.

\bibitem{begassat2019handel}
O.~B{\'e}gassat, B.~Kolad, N.~Gailly, and N.~Liochon, ``Handel: Practical multi-signature aggregation for large byzantine committees,'' \emph{arXiv preprint arXiv:1906.05132}, 2019.

\bibitem{li2020gosig}
P.~Li, G.~Wang, X.~Chen, F.~Long, and W.~Xu, ``Gosig: a scalable and high-performance byzantine consensus for consortium blockchains,'' in \emph{Proceedings of the 11th ACM Symposium on Cloud Computing}, 2020, pp. 223--237.

\bibitem{yin2019hotstuff}
M.~Yin, D.~Malkhi, M.~K. Reiter, G.~G. Gueta, and I.~Abraham, ``Hotstuff: {BFT} consensus with linearity and responsiveness,'' in \emph{Proceedings of the 2019 ACM Symposium on Principles of Distributed Computing}, 2019, pp. 347--356.

\bibitem{nakamoto2008bitcoin}
S.~Nakamoto, ``Bitcoin: A peer-to-peer electronic cash system,'' Tech. Rep., 2008.

\bibitem{meng2018committee}
Y.~Meng, Z.~Cao, and D.~Qu, ``A committee-based byzantine consensus protocol for blockchain,'' in \emph{2018 IEEE 9th International Conference on Software Engineering and Service Science (ICSESS)}.\hskip 1em plus 0.5em minus 0.4em\relax IEEE, 2018, pp. 1--6.

\bibitem{kwon2014tendermint}
J.~Kwon, ``Tendermint: Consensus without mining,'' \emph{Draft v. 0.6, fall}, vol.~1, no.~11, 2014.

\bibitem{baudet2019LibraBFT}
M.~Baudet, A.~Ching, A.~Chursin, G.~Danezis, F.~Garillot, Z.~Li, D.~Malkhi, O.~Naor, D.~Perelman, and A.~Sonnino, ``State machine replication in the {L}ibra blockchain,'' \emph{The Libra Assn., Tech. Rep}, 2019.

\bibitem{daian2019snow}
P.~Daian, R.~Pass, and E.~Shi, ``Snow white: Robustly reconfigurable consensus and applications to provably secure proof of stake,'' in \emph{International Conference on Financial Cryptography and Data Security}.\hskip 1em plus 0.5em minus 0.4em\relax Springer, 2019, pp. 23--41.

\bibitem{hanke2018dfinity}
T.~Hanke, M.~Movahedi, and D.~Williams, ``Dfinity technology overview series, consensus system,'' \emph{arXiv preprint arXiv:1805.04548}, 2018.

\bibitem{castro1999practical}
M.~Castro, B.~Liskov \emph{et~al.}, ``Practical byzantine fault tolerance,'' in \emph{OsDI}, vol.~99, no. 1999, 1999, pp. 173--186.

\bibitem{lamport2019byzantine}
L.~Lamport, R.~Shostak, and M.~Pease, ``The byzantine generals problem,'' in \emph{Concurrency: the works of Leslie Lamport}, 2019, pp. 203--226.

\bibitem{amoussou2020rational}
Y.~Amoussou-Guenou, B.~Biais, M.~Potop-Butucaru, and S.~Tucci-Piergiovanni, ``Rational behavior in committee-based blockchains,'' \emph{Cryptology ePrint Archive}, 2020.

\bibitem{hotstuffArxiv}
M.~Yin, D.~Malkhi, M.~K. Reiter, G.~G. Gueta, and I.~Abraham, ``Hotstuff: Bft consensus in the lens of blockchain,'' 2019.

\bibitem{buterin2017casper}
V.~Buterin and V.~Griffith, ``Casper the friendly finality gadget,'' \emph{arXiv preprint arXiv:1710.09437}, 2017.

\bibitem{abraham2021rotating}
I.~Abraham, K.~Nayak, and N.~Shrestha, ``Optimal good-case latency for rotating leader synchronous bft,'' \emph{Cryptology ePrint Archive}, 2021.

\bibitem{giridharan2023beegees}
N.~Giridharan, F.~Suri-Payer, M.~Ding, H.~Howard, I.~Abraham, and N.~Crooks, ``Beegees: stayin'alive in chained bft,'' in \emph{Proceedings of the 2023 ACM Symposium on Principles of Distributed Computing}, 2023, pp. 233--243.

\bibitem{gst}
\BIBentryALTinterwordspacing
C.~Dwork, N.~Lynch, and L.~Stockmeyer, ``Consensus in the presence of partial synchrony,'' \emph{J. ACM}, vol.~35, no.~2, pp. 288--323, apr 1988. [Online]. Available: \url{https://doi.org/10.1145/42282.42283}
\BIBentrySTDinterwordspacing

\bibitem{boneh2003aggregate}
D.~Boneh, C.~Gentry, B.~Lynn, and H.~Shacham, ``Aggregate and verifiably encrypted signatures from bilinear maps,'' in \emph{International conference on the theory and applications of cryptographic techniques}.\hskip 1em plus 0.5em minus 0.4em\relax Springer, 2003, pp. 416--432.

\bibitem{boneh2004short}
D.~Boneh, B.~Lynn, and H.~Shacham, ``Short signatures from the {W}eil pairing,'' \emph{Journal of cryptology}, vol.~17, no.~4, pp. 297--319, 2004.

\bibitem{coron2003boneh}
J.-S. Coron and D.~Naccache, ``Boneh et al.’sk-element aggregate extraction assumption is equivalent to the diffie-hellman assumption,'' in \emph{Advances in Cryptology-ASIACRYPT 2003: 9th International Conference on the Theory and Application of Cryptology and Information Security, Taipei, Taiwan, November 30--December 4, 2003. Proceedings 9}.\hskip 1em plus 0.5em minus 0.4em\relax Springer, 2003, pp. 392--397.

\bibitem{micali1999verifiable}
S.~Micali, M.~Rabin, and S.~Vadhan, ``Verifiable random functions,'' in \emph{40th annual symposium on foundations of computer science (cat. No. 99CB37039)}.\hskip 1em plus 0.5em minus 0.4em\relax IEEE, 1999, pp. 120--130.

\bibitem{abraham2016solidus}
I.~Abraham, D.~Malkhi, K.~Nayak, L.~Ren, and A.~Spiegelman, ``Solidus: An incentive-compatible cryptocurrency based on permissionless byzantine consensus,'' \emph{CoRR, abs/1612.02916}, 2016.

\bibitem{selfishmining}
I.~Eyal and E.~G. Sirer, ``Majority is not enough: Bitcoin mining is vulnerable,'' \emph{Communications of the ACM}, vol.~61, no.~7, pp. 95--102, 2018.

\bibitem{appliedHotstuff}
H.~Gogada, H.~Meling, L.~Jehl, and J.~I. Olsen, ``An extensible framework for implementing and validating byzantine fault-tolerant protocols,'' in \emph{Proceedings of the 5th Workshop on Advanced Tools, Programming Languages, and PLatforms for Implementing and Evaluating Algorithms for Distributed Systems}, ser. ApPLIED 2023.\hskip 1em plus 0.5em minus 0.4em\relax New York, NY, USA: Association for Computing Machinery, 2023.

\end{thebibliography}
